\documentclass[11pt,article]{elsarticle}

\usepackage{lineno,hyperref}
\usepackage[lmargin=.75in,rmargin=.75in,tmargin=1.in,bmargin=1in]{geometry}
\usepackage{amsmath}
\usepackage{amsthm}
\usepackage{amsfonts}
\usepackage{amssymb}
\usepackage{graphicx}
\usepackage{multicol}
\usepackage{wrapfig}
\usepackage{setspace}
\usepackage{epstopdf}
\usepackage{graphicx}
\usepackage{caption}
\usepackage{subcaption}
\usepackage[dvipsnames]{xcolor}

\usepackage{algorithm}
\usepackage[noend]{algpseudocode}
\newtheorem*{remark}{Remark}

\DeclareMathOperator{\sign}{sign}

\journal{arXiv}

\title{Modeling metallic fatigue data using the Birnbaum--Saunders distribution}

\begin{document}
%\maketitle
\begin{frontmatter}

\author[kfupm]{Zaid Sawlan \corref{corrauthor}}
\cortext[corrauthor]{Corresponding author}
\ead{zaid.sawlan@kfupm.edu.sa}
\author[Ur]{Marco Scavino}
\ead{marco.scavino@fcea.edu.uy}
\author[KAUST,RWTH]{Ra\'ul Tempone}
\ead{raul.tempone@kaust.edu.sa}

\address[kfupm]{Department of Mathematics, King Fahd University of Petroleum and Minerals, Dhahran, Saudi Arabia}
\address[Ur]{Instituto de Estad\'{\i}stica (IESTA), Universidad de la Rep\'ublica, Montevideo, Uruguay}
\address[KAUST]{CEMSE, King Abdullah University of Science and Technology, Thuwal, Saudi Arabia}
\address[RWTH]{Chair of Mathematics for Uncertainty Quantification, RWTH Aachen University, Aachen, Germany}

\begin{abstract}
%This work employs the Birnbaum--Saunders distribution to model metallic fatigue and compares its performance to fatigue-limit models based on the normal distribution. First, we fit data for 85 fatigue experiments with constant amplitude cyclic loading applied to unnotched sheet specimens of 75S-T6 aluminum alloys. The fit obtained by the Birnbaum--Saunders distribution is noticeably better than the normal distribution. Then, we define new equivalent stress for two fatigue experiment types: tension-compression and tension-tension. With the new equivalent stress, the statistical fit improves for both distributions, with a slight preference for the Birnbaum--Saunders distribution. In addition, we analyze a dataset of rotating-bending fatigue experiments applied to 101 round bar specimens of 75S-T6 aluminum. Finally, we consider a well-known dataset of bending tests applied to 125 specimens of carbon laminate. Overall, the Birnbaum--Saunders distribution provides better fit results under fatigue-limit models with various experimental setups.
This work employs the Birnbaum–Saunders distribution to model the fatigue life of metallic materials under cyclic loading and compares it with the normal distribution. Fatigue-limit models are fitted to three datasets of unnotched specimens of 75S-T6 aluminum alloys and carbon laminate with different loading types. A new equivalent stress definition that accounts for the effect of the experiment type is proposed. The results show that the Birnbaum–Saunders distribution consistently outperforms the normal distribution in fitting the fatigue data and provides more accurate predictions of fatigue life and survival probability.
\end{abstract}


\end{frontmatter}

%\linenumbers

\section{Introduction}

% Motivation and general ideas
Fatigue-life prediction is vital to preventing the failure of mechanical parts under cyclic loading. Stress-life models, or S-N curves, are usually used to model fatigue life \cite{schijve, schijvebook, fatemi}. Although many models relate stress to the fatigue life, with probabilistic models, the fatigue life is often assumed to be a log-normal random variable \cite{fatigue1, pasmee, rflm, ryan} or to follow the Weibull distribution \cite{fatigue1, weibull}. A comprehensive review of plausible models for fatigue life is available in \cite{tridello2022}. For a specified stress-life model, a corresponding fatigue strength model can be induced and then used to make predictions \cite{meeker2022modern}.

% goal of the study
This work considers Birnbaum--Saunders distributions \cite{bs1} that were introduced to model fatigue failure time under cyclic loading. We aim to study the use of Birnbaum--Saunders distributions and analyze their fit results compared with the dominant choice of normal/log-normal distributions. Thus, we use two datasets of fatigue experiments applied to specimens of 75S-T6 aluminum alloys \cite{gbj, hlg} and a dataset of bending tests of carbon laminate \cite{shimokawa1987}. Moreover, we study the effect of equivalent stress model when fatigue data includes different types of experiments.

% literature review of BS distributions
Birnbaum--Saunders distributions were introduced as a two-parameter family of life distributions \cite{bs1}. Several studies have used these distributions to fit fatigue datasets using the maximum likelihood (ML) \cite{birnbaum1969estimation, inf_bs} and Bayesian methods \cite{achcar1993inferences, tsionas2001bayesian, xu2011bayesian}. In addition, many variations have been proposed, such as the log-linear model for the Birnbaum--Saunders distribution \cite{rieck1991log} and bivariate log Birnbaum--Saunders distribution \cite{kundu2015bivariate}. An extensive review of the Birnbaum--Saunders distribution and its generalizations is provided in \cite{leiva2015birnbaum,balakrishnan}.

% goal of the study
%We aim to study the use of Birnbaum--Saunders distributions and analyze their fit results compared with the dominant choice of normal distributions. Thus, we use two datasets of fatigue experiments applied to specimens of 75S-T6 aluminum alloys \cite{gbj, hlg} and a dataset of bending tests of carbon laminate \cite{shimokawa1987}. 

% fatigue-limit models and modeling log(N)
For the S-N models, many possible regression models could be considered. We focus only on the fatigue-limit models with constant and nonconstant variance (or the shape parameter). The fatigue-life variable is modeled by the log-normal distribution or Birnbaum--Saunders distribution. Equivalently, the logarithm of $N$ is modeled by the normal and sinh-normal distributions, respectively. However, we demonstrate that modeling $log(N)$ as a Birnbaum--Saunders distribution improves the fit results. This proposed model is unprecedented in the literature, to the best of our knowledge. 

% Datasets
The first dataset is the same data considered in \cite{fatigue1}, where fatigue-limit models and random fatigue-limit models were calibrated using the normal and Weibull distributions. We recalibrate fatigue-limit models using the Birnbaum--Saunders distribution. In addition, we propose a new equivalent stress model that accommodates different experiment types in Dataset~1. For Dataset~2, fatigue data corresponds to rotating-bending experiments applied to round bar specimens with different minimum-section diameters \cite{hlg}. Again, we calibrate fatigue-limit models and compare the fit of the normal and Birnbaum--Saunders distributions. As Dataset~3, we use the laminate panel data \cite{shimokawa1987, pasmee}, and calibrate and compare our proposed models.

% Main Findings
The results show that modeling $log(N)$ by mean of Birnbaum--Saunders distribution improves fitting systematically in all three datasets and using different variations of fatigue-limit models. It is also expected that such a choice would improve fitting with different S-N models. However, it is not our goal to find the best model for each dataset. For Dataset~1, our proposed equivalent stress also improves data fitting using different models and distributions. 

% paper structure
This paper is organized as follows. Section~\ref{sec2} considers the fatigue data for unnotched sheet specimens of 75S-T6 aluminum alloys. Six variations of fatigue-limit models are introduced in Section~\ref{sec2}, comparing stress-life models fitted to the data using the normal and Birnbaum--Saunders distributions. Next, Section~\ref{Sec-Seq} proposes a new equivalent stress definition to eliminate the effect of the experiment type. Then, Section~\ref{sec4} presents fatigue data corresponding to the unnotched round bar specimens of different sizes, followed by calibration and model comparison. Then, laminate panel data are fitted and analyzed in Section~\ref{sec5} using the predefined fatigue-limit models. Finally, the conclusions are presented in Section~\ref{sec6}.

\section{Model calibration and comparison for Dataset 1}
\label{sec2}

\subsection{Description of Dataset 1}
\label{Data1}
Dataset~1 consists of 85 fatigue experiments that applied constant amplitude cyclic loading to unnotched sheet specimens of 75S-T6 aluminum alloys \cite[Table~3, pp. 22--24]{gbj}. The following data are recorded for each specimen:
\begin{itemize}
\item maximum stress, $S_{max}$, measured in ksi units;
\item cycle ratio, $R$, defined as the minimum to maximum stress ratio. The ratio $R$ is positive when the experiment corresponds to tension-tension loading and negative when the experiment corresponds to tension-compression loading;
\item fatigue life, $N$, defined as the number of load cycles at which fatigue failure occurred; and
\item a binary variable (0/1) to denote whether the test stopped before failure (run-out).
\end{itemize}
In 12 of the 85 experiments, the specimens remained unbroken when the tests were stopped. 
%The specimen sheets have dog-bone shape with thickness $0.09$ inches as illustrated in Figure \ref{specimen_data1}.

% \begin{figure}[h!]
% \centering
% \includegraphics[width=14cm]{unnotched.eps}
% \caption{Dataset 1: shape of a sheet specimen where the cross-sectional thickness is $0.09$ inches.}
% \label{specimen_data1}
% \end{figure} 
%The recorded number of load cycles for these 12 experiments is the lower bound of an interval in which failure would have occurred had the test been continued. If specimens buckled or failed outside the test section, they are not included in the dataset.

\subsection{Fatigue-limit models}
\label{FLmodels}
The fatigue life should be modeled for a stress quantity defined for any cycle ratio. Therefore, we use Walker's model to define the equivalent stress:
\begin{equation}
    S_{eq} = S_{max}(1-R)^q,
\end{equation} 
where $q$ is a fitting parameter. In the upcoming sections, we consider fatigue-limit models where the location parameter is given by $A_1 + A_2 log_{10}(S_{eq} - A_3)$ and the fatigue-limit parameter $A_3$ is a threshold parameter where fatigue life becomes infinite when the equivalent stress is lower than $A_3$. Multiple fatigue-limit models could be created based on the choice of the distribution of the fatigue life, $N$. We consider three choices as follows.

\subsubsection{Model Ia}
%\label{model1} 
In Model Ia, we assume fatigue life is modeled using a log-normal distribution, or equivalently, that $\log_{10}(N)$ is modeled with a normal distribution with a mean of 
\begin{equation}
\mu(S_{eq}) = A_1 + A_2 \, \log_{10}( S_{eq} - A_3)\,,\:\:\textrm{if}\:\:S_{eq} > A_3  \label{mu}
\end{equation}
and a constant standard deviation of $\sigma(S_{eq}) = \tau$. Moreover, fatigue experiments are assumed independent, and run-outs are modeled using the survival probability. Thus, the likelihood function for Model Ia is given by
\begin{equation}
\label{likelihood}
L(A_1, A_2, A_3, \tau, q; \mathbf{n}) = \prod_{i=1}^{m} \left[ \frac{1}{n_i \log(10)} g(\log_{10}(n_i)\,;\mu(S_{eq})\,,\tau) \right]^{\delta_i}  \, 
\left[ 1- \Phi \left( \frac{\log_{10}(n_i) - \mu(S_{eq})}{\tau} \right) \right]^{1 - \delta_i} \,,
\end{equation}
where $g(t; \mu, \sigma) = \frac{1}{\sqrt{2 \pi} \,\sigma} exp \left\{ - \frac{(t - \mu)^2}{2 \sigma^2} \right\}\,,$ $\Phi$ is the cumulative distribution function of the standard normal distribution, and
\begin{equation*}
\delta_i = \left\{
\begin{array}{rl}
1 & \text{if } n_i \text{ is a failure}\\
0 & \text{if } n_i \text{ is a run-out\,.}
\end{array} \right.
\end{equation*} 

\begin{remark}
Model Ia and the likelihood function \eqref{likelihood} have been used in \textit{[Babuška, Ivo, et al. "Bayesian inference and model comparison for metallic fatigue data." Computer Methods in Applied Mechanics and Engineering 304 (2016): 171-196.]}    
\end{remark}

\subsubsection{Model IIa}
For Model~IIa, we assume fatigue life is modeled using the Birnbaum--Saunders distribution, or equivalently, that $\log(N)$ is modeled with a sinh-normal distribution \cite{balakrishnan} with a constant shape parameter $\alpha$, a scale parameter of 2, and a location parameter $\mu(S_{eq})$ given by \eqref{mu}. Under this assumption, the likelihood function for Model~IIa is given by
\begin{equation*}
L(A_1, A_2, A_3, \alpha, q; \mathbf{n}) = \prod_{i=1}^{m} \left[ \frac{1}{n_i} h(\log(n_i)\,;\alpha , \mu(S_{eq})) \right]^{\delta_i}  \, 
\left[ 1- \Phi \left( \frac{2}{\alpha} \sinh\left( \frac{\log(n_i) -\mu(S_{eq})}{2} \right) \right) \right]^{1 - \delta_i} \,,
\end{equation*}
where $h(y ; \alpha, \mu) = \frac{1}{\alpha \sqrt{2 \pi}} \cosh\left(\frac{y-\mu}{2}\right) \exp\Big( - \frac{2}{\alpha^2} \sinh^2\left(\frac{y-\mu}{2}\right)\Big) \,,y >0, \textrm{ and } \alpha, \mu >0 \,.$

\subsubsection{Model IIIa}
% \label{model2}
For Model~IIIa, we assume $\log_{10}(N)$ is modeled using a Birnbaum--Saunders distribution with a constant shape parameter $\alpha$ and a location parameter $\mu(S_{eq})$, given by \eqref{mu}. The resulting distribution for $N$ is not the so-called log Birnbaum--Saunders distribution reported in \cite{balakrishnan}. However, the distribution of $N$ is obtained similarly to deriving the log-normal distribution.
\begin{equation*}
L(A_1, A_2, A_3, \alpha, q; \mathbf{n}) = \prod_{i=1}^{m} \left[ \frac{1}{n_i \log(10)} k(\log_{10}(n_i)\,;\alpha , \mu(S_{eq})) \right]^{\delta_i}  \, 
\left[ 1- \Phi \left( \frac{1}{\alpha} \left( \sqrt{\frac{\log_{10}(n_i)}{\mu(S_{eq})}} - \sqrt{\frac{\mu(S_{eq})}{\log_{10}(n_i)}} \right) \right) \right]^{1 - \delta_i} \,,
\end{equation*}
where $k(y ; \alpha, \mu) = \frac{1}{\sqrt{2 \pi}} \frac{(y + \mu)}{2 \alpha \sqrt{\mu} y^{3/2}} \exp \Big\{ - \frac{1}{2 \alpha^2} \left( \frac{y}{\mu} +  \frac{\mu}{y} - 2 \right) \Big\} \,, y >0, \textrm{ and } \alpha, \mu >0 \,.$

The three models are now fitted to Dataset~1 by maximizing the mentioned likelihood functions. Numerically, only the log-likelihood can be evaluated, and we maximize the log-likelihood instead. The ML estimates (MLEs) and the maximum log-likelihood value are reported in Table~\ref{mle1a}. The estimated parameters for Model~IIa have a different scale compared with those in Models~Ia and IIIa. The change in scale is only because Model~IIa models $log(N)$ instead of $log_{10}(N)$. However, all applied likelihood functions are normalized; therefore, the performance of the fit is not affected by the logarithm base selection.   

\begin{table}[h!]
\begin{center}
\caption{Maximum likelihood estimates for Models~Ia, IIa, and IIIa}
\begin{tabular}{|c|c|c|c|c|c|c|c|}
\hline
 & $A_1$ & $A_2$ & $A_3$ & $q$ & $\tau / \alpha$ & Max log-likelihood \\
\hline
Model Ia & 7.38 &   -2.01 &  35.04 &  0.5628 &   0.5274 & -950.16 \\
\hline
Model IIa & 18.81 &  -5.68  &  33.11 &  0.5390  &  1.54  & -960.68 \\
\hline
Model IIIa & 7.22 &   -1.90 &  35.32 &  0.5574 &   0.0933 & -938.90 \\
\hline
\end{tabular}
\label{mle1a}
\end{center}
\end{table}

\begin{figure}[h!]
\includegraphics[width=18cm]{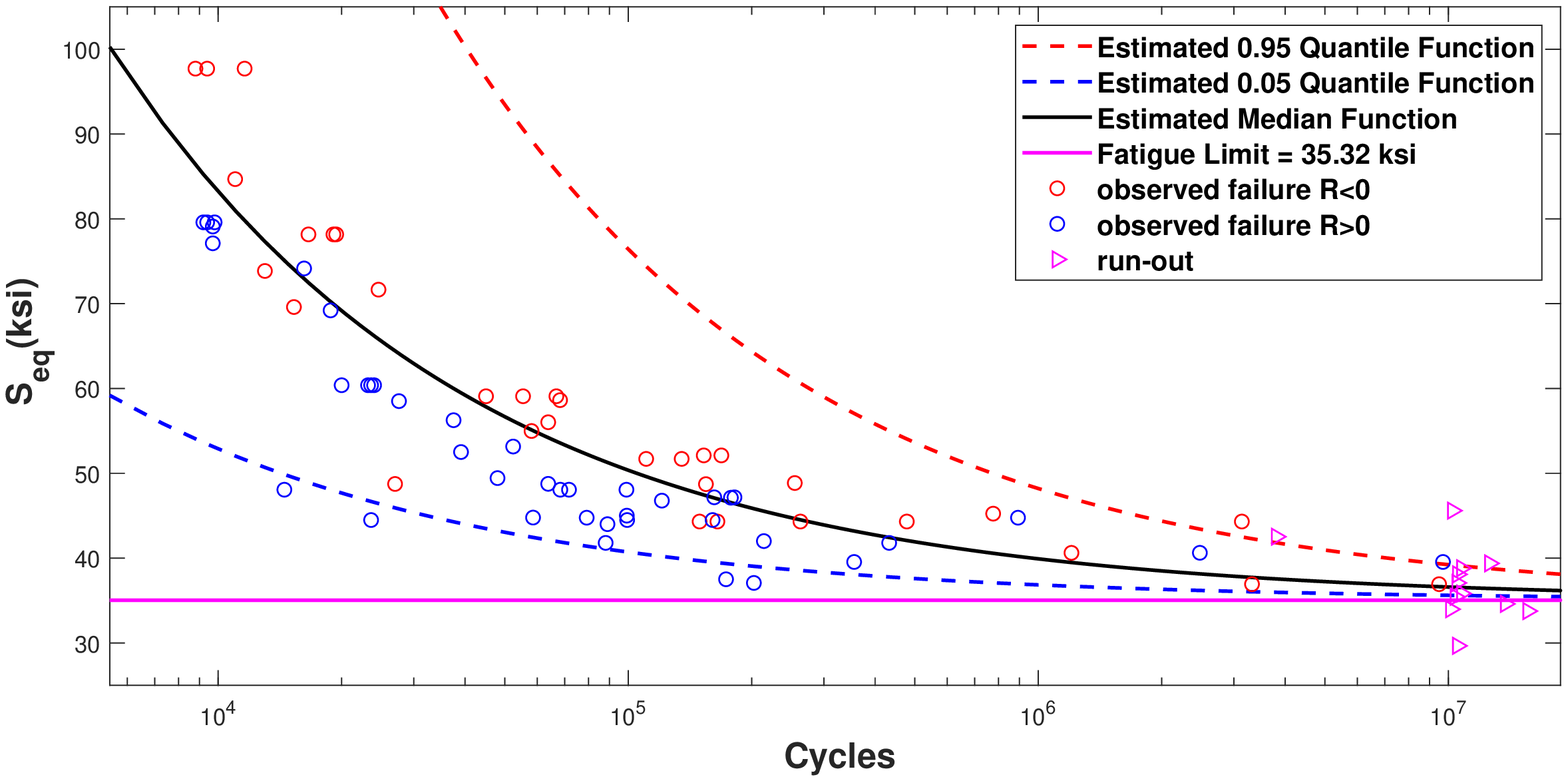}
\caption{Model Ia: $\log_{10}(N) \sim N(\mu(S_{eq}), \sigma)$ and $S_{eq} = S_{max}(1-R)^q$.}
\label{fit1}
\end{figure}

\begin{figure}[h!]
\includegraphics[width=18cm]{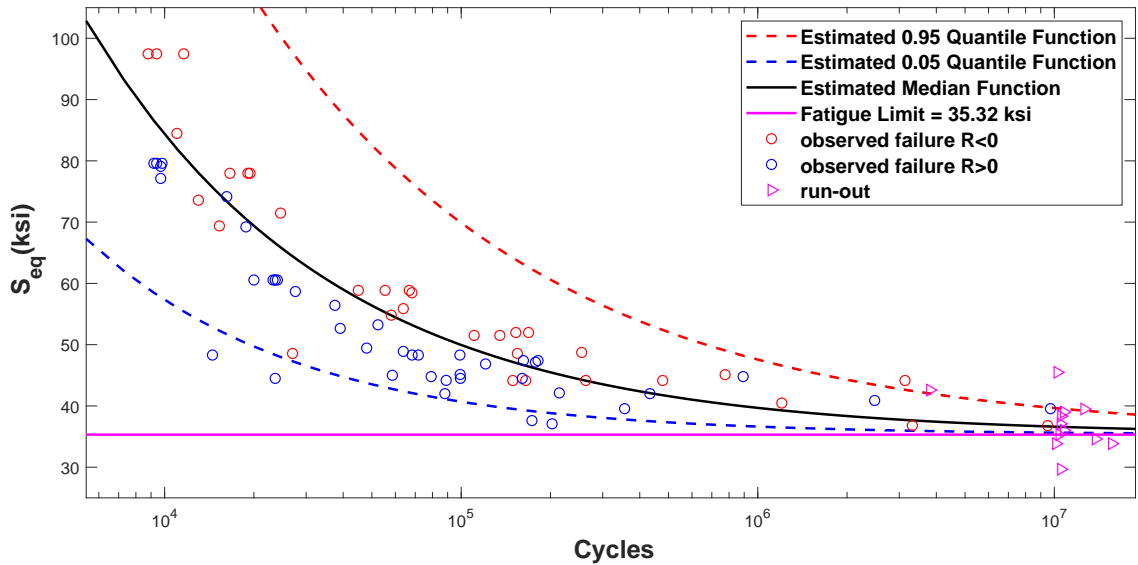}
\caption{Model IIIa: $\log_{10}(N) \sim BS(\alpha, \mu(S_{eq}))$ and $S_{eq} = S_{max}(1-R)^q$.}
\label{fit2}
\end{figure}

The results in Table~\ref{mle1a} indicate that Models~Ia and IIIa provide the best fit for Dataset~1. We visualize the fit using the 0.05 and 0.95 quantile functions and the median function. The data can also be plotted given the MLE of $q$. We distinguish data based on the experiment type or stress ratio. Figures~\ref{fit1} and \ref{fit2} illustrate the quantile functions of Models~Ia and IIIa obtained using the MLE parameters. The quantile functions of Model~IIIa produce a better fit than those of Model~Ia, which coincides with the fact that Model~IIIIa has the highest log-likelihood value among the three models in Table~\ref{mle1a}. We also observe that data seem segregated by the median according to the experiment type: tension-tension ($R>0$) and tension-compression ($R<0$). We analyze this behavior further in Section~\ref{Sec-Seq}. 

We allow the standard deviation or shape parameter to be nonconstant to improve the fit of the three previous models. In particular, we assume this parameter is a function of the equivalent stress. With the same probability distributions previously considered, we introduce three new fatigue-limit models with nonconstant standard deviation/shape parameters.

\subsubsection{Model Ib}
Analogous to Model~Ia, for Model~Ib, we assume $\log_{10}(N)$ has a normal distribution with the mean function $\mu(S_{eq})$ defined in \ref{mu}. However, the standard deviation is assumed nonconstant and given by $\sigma(S_{eq}) = 10^{(B_1 + B_2 \log_{10}(S_{eq}))}$. The resulting likelihood function is equivalent to that derived for Model~Ia. 

\subsubsection{Model IIb}
For Model~IIb, the fatigue life $N$ is modeled by the Birnbaum--Saunders distribution with location parameter $\mu(S_{eq})$ (defined in \ref{mu}) and nonconstant shape parameter $\alpha(S_{eq}) = 10^{(B_1 + B_2 \log_{10}(S_{eq}))}$.

\subsubsection{Model IIIb}
For Model~IIIb, $\log_{10}(N)$ is modeled using the Birnbaum--Saunders distribution with the location parameter $\mu(S_{eq})$ and nonconstant shape parameter $\alpha(S_{eq}) = 10^{(B_1 + B_2 \log_{10}(S_{eq}))}$.

The new Models~Ib, IIb, and IIIb are calibrated to fit Dataset~1, and the MLEs of the parameters of these models are presented in Table~\ref{mle1b}. Comparing the maximum log-likelihood values in Tables~\ref{mle1a} and \ref{mle1b} reveals that the fit improved considerably for all models. In contrast, the difference between the new models decreased, with Model IIIb still providing the best fit.

\begin{table}[h!]
\begin{center}
\caption{Maximum likelihood estimates for Models~Ib, IIb, and IIIb}
\begin{tabular}{|c|c|c|c|c|c|c|c|c|}
\hline
 & $A_1$ & $A_2$ & $A_3$ & $q$ & $B_1$ &  $B_2$ & Max log-likelihood \\
\hline
Model Ib & 6.72 & -1.57 & 36.21 & 0.5510 &  4.55 & -2.89  &  -920.51  \\
\hline
Model IIb & 16.56  & -4.26  &  35.51  &  0.5239  &  5.54  &  -3.21 & -926.97  \\
\hline
Model IIIb & 6.70 & -1.56 & 36.24 & 0.5501 & 2.90 & -2.34  & -917.38 \\
\hline
\end{tabular}
\label{mle1b}
\end{center}
\end{table}

\begin{figure}[h!]
\includegraphics[width=18cm]{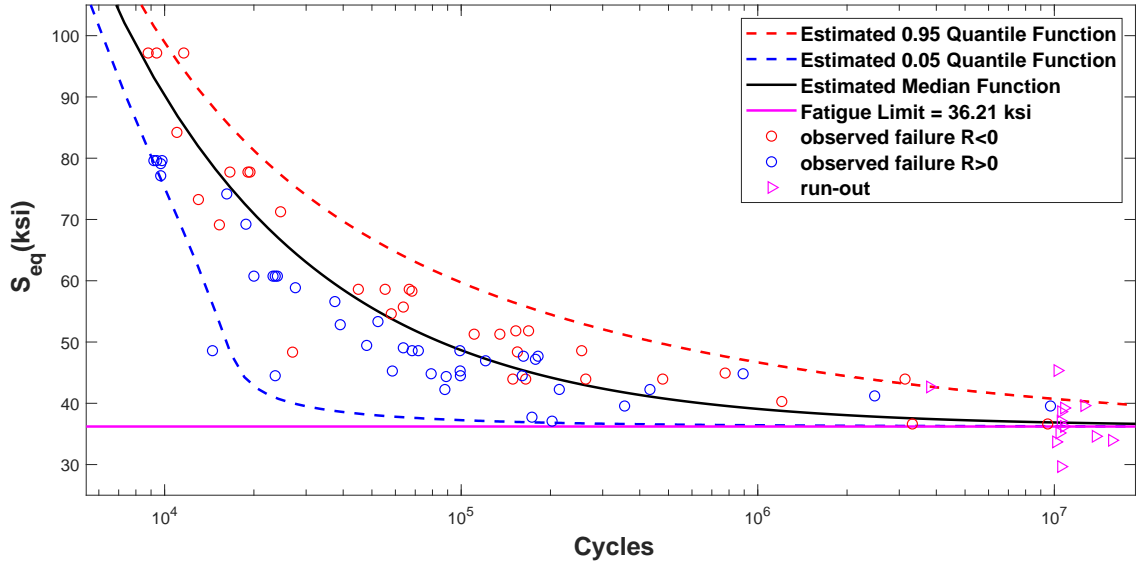}
\caption{Model Ib: $\log_{10}(N) \sim N(\mu(S_{eq}), \sigma(S_{eq}))$ and $S_{eq} = S_{max}(1-R)^q$.}
\label{fit1b}
\end{figure}

\begin{figure}[h!]
\includegraphics[width=18cm]{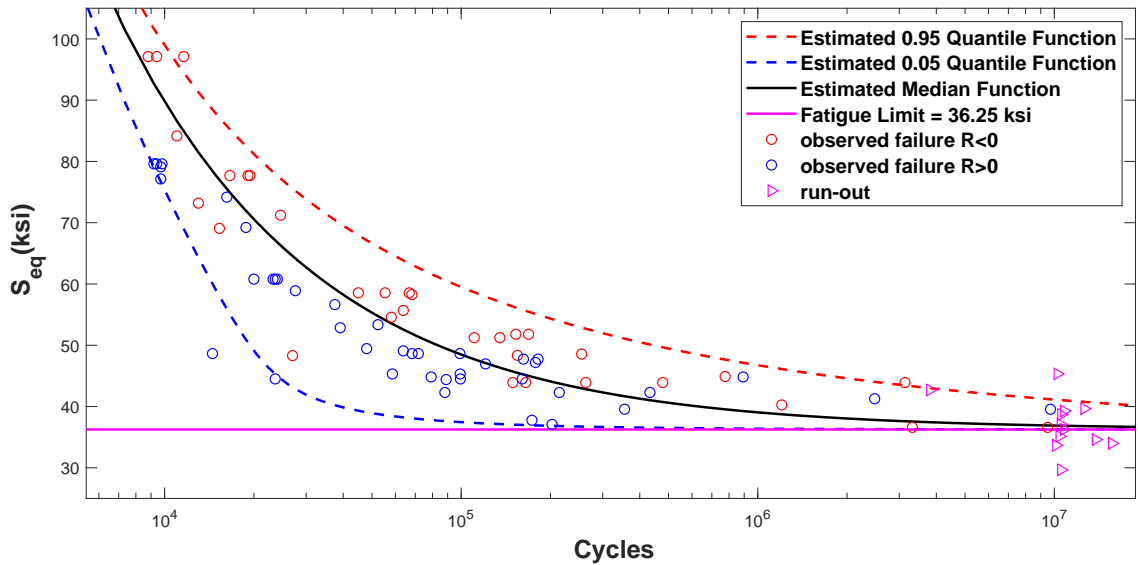}
\caption{Model IIIb: $\log_{10}(N) \sim BS(\alpha(S_{eq}), \mu(S_{eq}))$ and $S_{eq} = S_{max}(1-R)^q$.}
\label{fit2b}
\end{figure}

We compare the fit of Models~Ib and IIIb using the quantile functions in Figures~\ref{fit1b} and \ref{fit2b}. Both models produced significantly improved fit compared with Figures~\ref{fit1} and \ref{fit2}. The fatigue-limit parameter slightly increased, and the 0.05 quantile converges rapidly to its asymptote as the equivalent stress approaches the fatigue limit. In both cases, the data remain mostly partitioned by the median into the two experiment types.

\subsection{Model comparison}
%\label{sec3c}
Using a classical approach, we compute some popular information criteria, such as the Akaike information criterion (AIC) \cite{aic}, Bayesian information criterion (BIC) \cite{bic1,bic2}, and AIC with correction \cite{aicc}, which are based on the maximized log-likelihood values. Such measures consider the goodness of fit and complexity of the models regarding the number of parameters. Table~\ref{Capp} contains the maximum log-likelihood values corresponding to the models introduced in Section~\ref{FLmodels} and the classical information criterion computations. 

\begin{table}[h!]
\begin{center}
\caption{Classical information criteria.}
\begin{tabular}{|c|c|c|c|c|c|c|c|}
\hline
Models &  \bf{Ia} & \bf{Ib} & \bf{IIa} & \bf{IIb} & \bf{IIIa} & \bf{IIIb} \\
\hline
Maximum log-likelihood & -950.16 & -920.51 & -960.68  &   -926.97  & -938.90 & -917.38 \\
\hline
Akaike information criterion (AIC) & 1910.3 & 1853.0 &  1931.4 &  1865.9  & 1887.8  & 1846.8   \\
\hline
Bayesian information criterion (BIC) & 1922.5 & 1867.7 &  1943.6  & 1880.6    &  1900.0  &  1861.4   \\
\hline
Akaike information criterion with correction & 1911.1 & 1854.1 &  1932.1 &  1867.0    & 1888.5 &  1847.8  \\
\hline
\end{tabular}
\label{Capp}
\end{center}
\end{table}

\section{Analysis of the stress ratio effect and equivalent stress for Dataset~1}
\label{Sec-Seq}
In all previous models, the equivalent stress is based on Walker's model \cite{walker}, which is $S_{max}(1-R)^q$. The following analysis in Table~\ref{Walker} reveals that the parameter $q$ is related to the sign of the cycle ratio $R$.

\begin{table}[h!]
\begin{center}
\caption{Maximum likelihood estimates for Models I and II with $S_{eq}=S_{max}(1-R)^q$.}
\begin{tabular}{|c|c|c|c|c|c|c|c|c|}
\hline
Model & Data & $A_1$ & $A_2$ & $A_3$ & $q$ & $\tau / \alpha /  B_1$ &  $B_2$ & Max log-likelihood \\
\hline
Ia & $R<0$ & 7.94 & -2.10 & 61.43 & 1.37 &  0.3203 & --- & -403.56  \\
\hline
IIIa & $R<0$ & 7.89 & -2.10 & 57.643 & 1.2753 &  0.0575 & --- & -399.64  \\
\hline
Ib & $R<0$ & 7.44 & -1.86 & 52.56 & 1.12 &  5.23 & -3.09 & -392.86  \\
\hline
IIIb & $R<0$ & 7.41 & -1.85 & 52.06 & 1.0986 &  3.40 & -2.50  & -392.23 \\
\hline
Ia & $R>0$ & 6.93 & -1.84 & 34.86 & 0.6304 &  0.5561 & ---  & -531.21 \\
\hline
IIIa & $R>0$ & 6.75 & -1.71 & 35.18 & 0.6269 &  0.0974 & --- & -523.95 \\
\hline
Ib & $R>0$ & 6.84 & -1.75 & 36.55 & 0.5410 &  8.11 & -5.07 & -500.46 \\
\hline
IIIb & $R>0$ & 6.75 & -1.69 & 36.62 & 0.5388 &  6.27 & -4.41 & -499.15 \\
\hline
\end{tabular}
\label{Walker}
\end{center}
\end{table}

\begin{table}[h!]
\begin{center}
\caption{Maximum likelihood estimates for Models I and II with $S_{eq} = S_{max}(\frac{1-R}{2})^{1+q}$.}
\begin{tabular}{|c|c|c|c|c|c|c|c|c|}
\hline
Model & Data & $A_1$ & $A_2$ & $A_3$ & \bf{$q$} & $\tau / \alpha /  B_1$ &  $B_2$ & Max log-likelihood \\
\hline
Ia & $R<0$ & 7.08 & -2.11 & 23.80 & 0.3679 &  0.3203 & --- & -403.56  \\
\hline
IIIa & $R<0$ & 7.09 & -2.10 & 23.81 & 0.2754 &  0.0575 & --- & -399.64  \\
\hline
Ib & $R<0$ & 6.82 & -1.86 & 24.26 & 0.1156 & 4.19  & -3.09 & -392.86  \\
\hline
IIIb & $R<0$ & 6.80 & -1.85 & 24.31 & 0.0986 &  2.57 & -2.50  & -392.23 \\
\hline
Ia & $R>0$ & 6.59 & -1.84 & 22.52 & -0.3696 &  0.5561 & ---  & -531.21 \\
\hline
IIIa & $R>0$ & 6.43 & -1.71 & 22.78 & -0.3732 &  0.0974 & --- & -523.95 \\
\hline
Ib & $R>0$ & 6.56 & -1.75 & 25.12 & -0.4590 &  7.29 & -5.07 & -500.46 \\
\hline
IIIb & $R>0$ & 6.54 & -1.74 & 25.19 & -0.4607 &  5.68 & -4.49 & -499.13 \\
\hline
\end{tabular}
\label{TU2}
\end{center}
\end{table}

In Table~\ref{Walker}, the fatigue-limit parameter, $A_3$, has a different scale based on the stress ratio. We divided $1-R$ by 2 in the equivalent stress formula to solve this. However, the estimated values of $q$ do not change. Therefore, we defined the equivalent stress as $S_{max}(\frac{1-R}{2})^{1+q}$ or $S_{a}(\frac{1-R}{2})^q$, where $S_a$ denotes the stress amplitude. Then, we recalibrated the proposed models in Table~\ref{TU2}. The fatigue-limit parameter has the same scale for $R>0$ and $R<0$. In contrast, the estimated value of $q$ changes signs with $R$. Thus, it seems reasonable to propose the following equivalent stress:
\begin{equation}
\label{eqnew}
S_{eq} = S_{max}\left(\frac{1-R}{2}\right)^{1-\sign(R)q}.
\end{equation}

\begin{table}[h!]
\begin{center}
\caption{Maximum likelihood estimates for Models I and II with $S_{eq} = S_{max}(\frac{1-R}{2})^{1-\sign(R)q}$.}
\begin{tabular}{|c|c|c|c|c|c|c|c|c|}
\hline
Model & $A_1$ & $A_2$ & $A_3$ & \bf{$q$} & $\tau / \alpha /  B_1$ &  $B_2$ & Max log-likelihood & AIC \\
\hline
Ia & 6.99 & -2.09 & 23.72 & 0.4310 &  0.4797 & --- & -942.55 & 1895.1 \\
\hline
IIIa & 6.81 & -1.95 & 23.98 & 0.4336 &  0.0845 & --- & -930.85 & 1871.7 \\
\hline
Ib & 6.58 & -1.76 & 24.56 & 0.4433 & 5.97  & -4.23 & -899.36 & 1810.7 \\
\hline
IIIb & 6.54 & -1.74 & 24.62 & 0.4436 &  4.38 & -3.65  & -897.67 & 1807.3 \\
\hline
\end{tabular}
\label{Walker2}
\end{center}
\end{table}

\begin{figure}[h!]
\includegraphics[width=18cm]{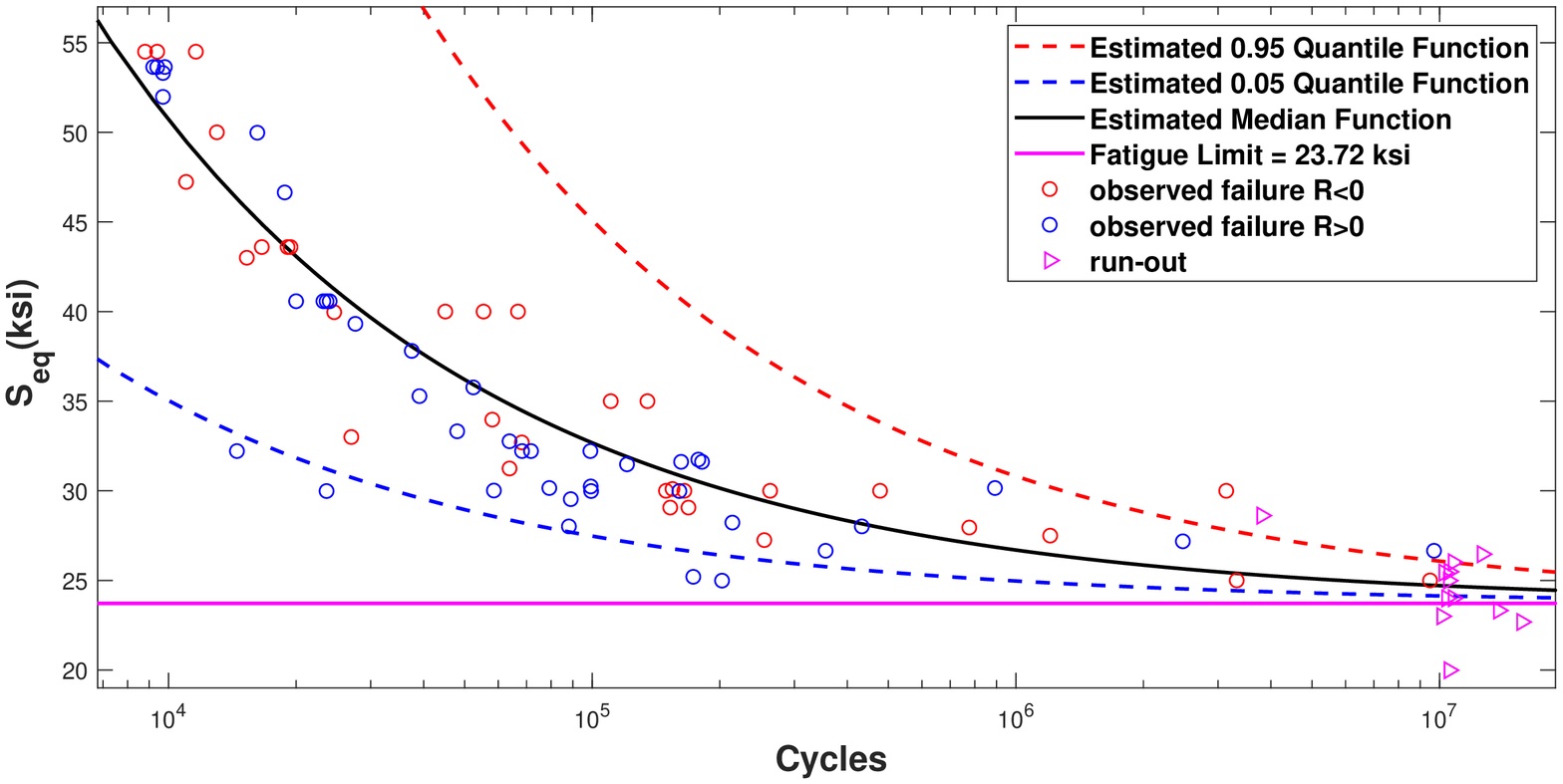}
\caption{Model Ia: $\log_{10}(N) \sim N(\mu(S_{eq}), \sigma)$ and $S_{eq} = S_{max}(\frac{1-R}{2})^{1-\sign(R)q}$.}
\label{fit1_Seq2}
\end{figure} 

\begin{figure}[h!]
\includegraphics[width=18cm]{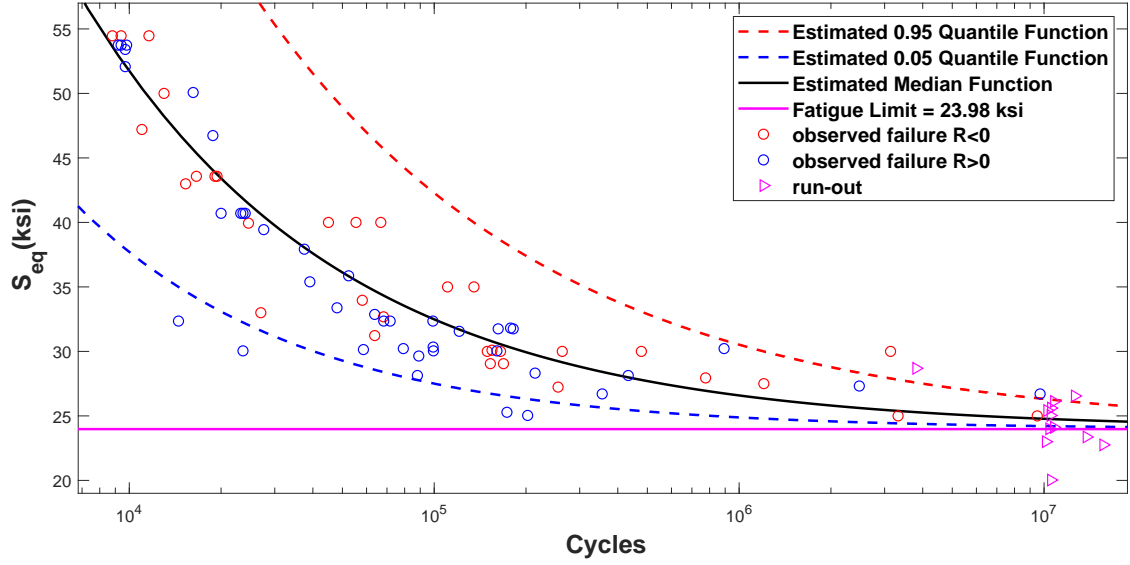}
\caption{Model IIIa: $\log_{10}(N) \sim BS(\alpha, \mu(S_{eq}))$ and $S_{eq} = S_{max}(\frac{1-R}{2})^{1-\sign(R)q}$.}
\label{fit2_Seq2}
\end{figure} 

Next, we calibrate the parameters using the full data ($R>0$ and $R<0$). Table~\ref{Walker2} presents the MLEs of Models Ia, IIIa, Ib, and IIIb, along with the maximum log-likelihood and AIC values. The fit is considerably improved in all cases using the equivalent stress \ref{eqnew}. 

\begin{figure}[h!]
\includegraphics[width=18cm]{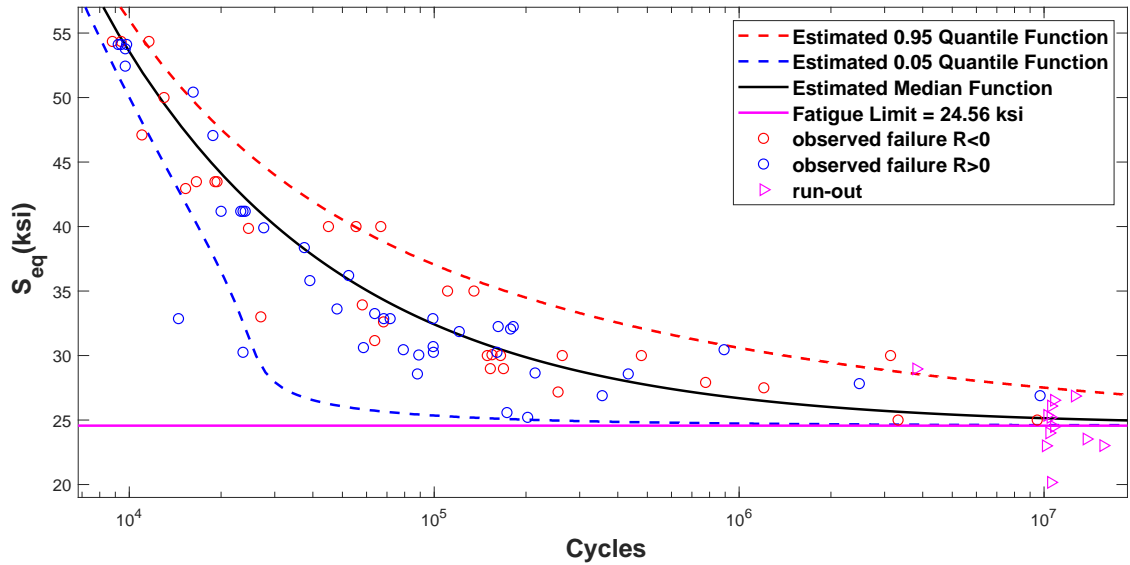}
\caption{Model Ib: $\log_{10}(N) \sim N(\mu(S_{eq}), \sigma(S_{eq}))$ and $S_{eq} = S_{max}(\frac{1-R}{2})^{1-\sign(R)q}$.}
\label{fit1b_Seq2}
\end{figure} 

\begin{figure}[h!]
\includegraphics[width=18cm]{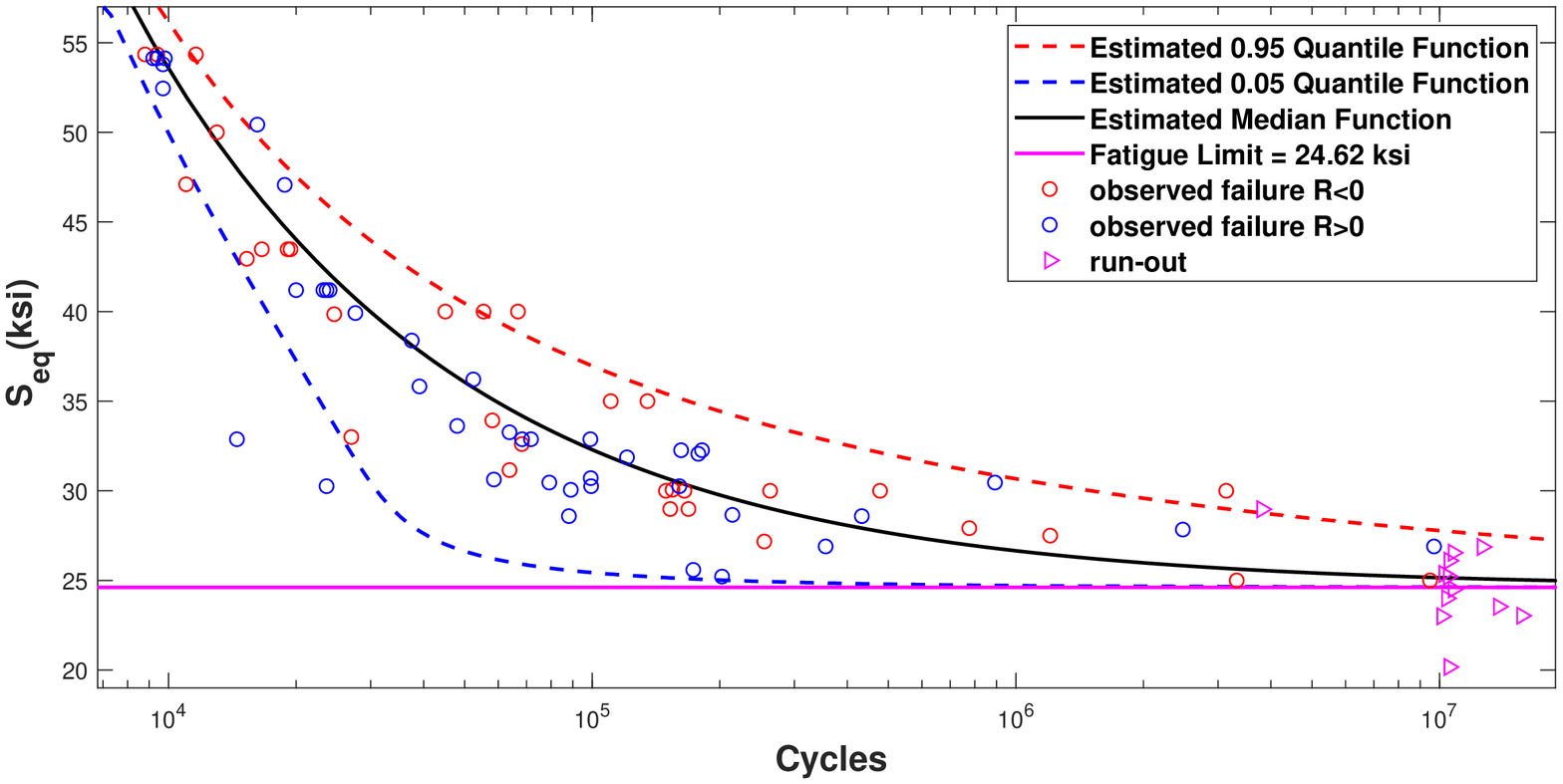}
\caption{Model IIIb: $\log_{10}(N) \sim BS(\alpha(S_{eq}), \mu(S_{eq}))$ and $S_{eq} = S_{max}(\frac{1-R}{2})^{1-\sign(R)q}$.}
\label{fit2b_Seq2}
\end{figure} 

Figures~\ref{fit1_Seq2} and \ref{fit2_Seq2} illustrate the new quantile functions of Models~Ia and IIIa with improved equivalent stress \ref{eqnew}. The variance is reduced compared to quantiles in Figures~\ref{fit1} and \ref{fit2}. In addition, the two data types are well distributed around the median. Furthermore, the fit can be slightly improved by adapting Huang's model \cite{huang} for $R>0.5$.

\subsection{Profile likelihood}

\begin{figure}[h!]
\centering
\begin{minipage}{.45\textwidth}
  \centering
  \includegraphics[width=9cm]{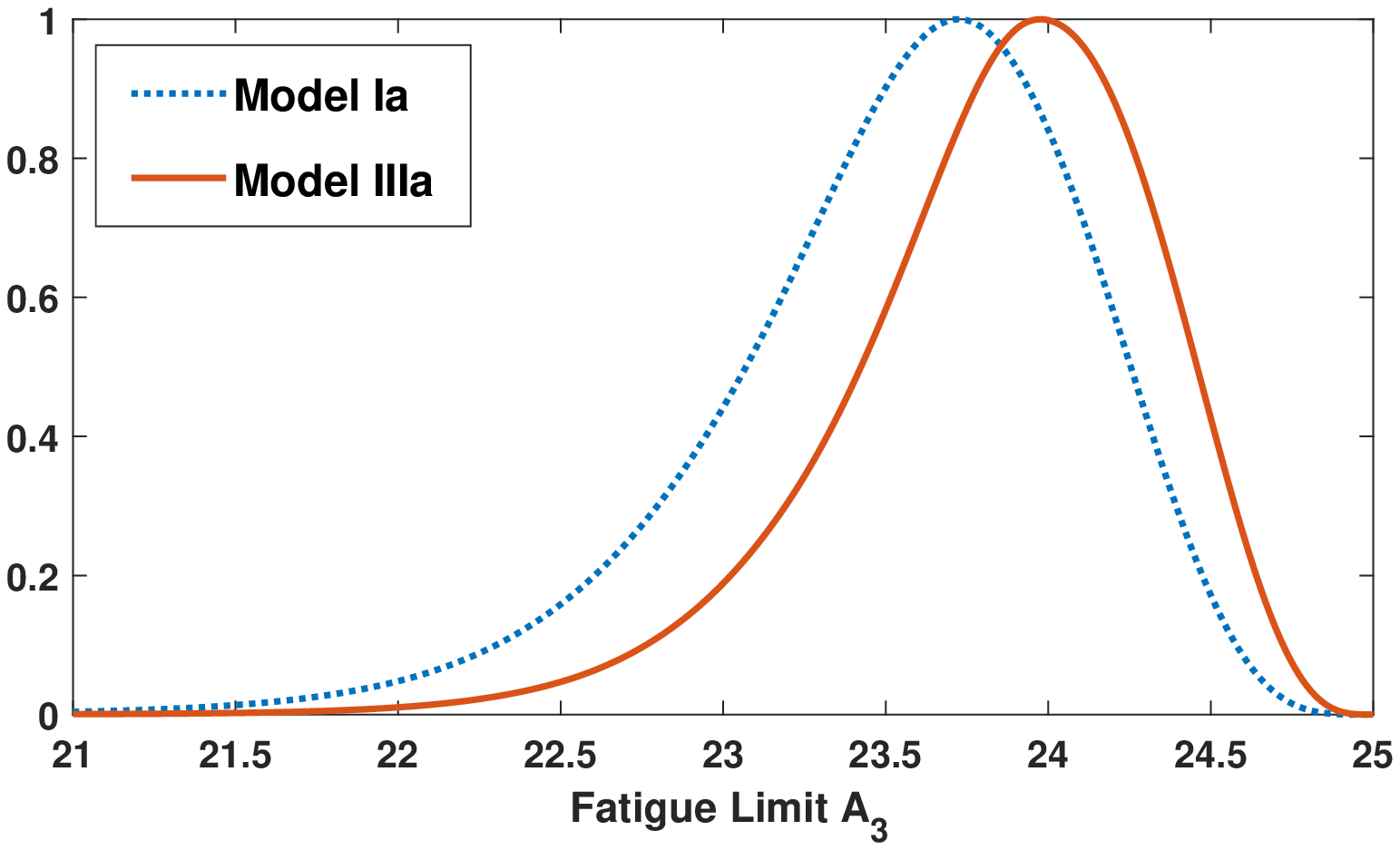}
  \captionof{figure}{Profile likelihoods of the fatigue-limit parameters using Models~Ia and IIIa.}
  \label{prof_like1}
\end{minipage}
~
\begin{minipage}{.45\textwidth}
\centering
  \includegraphics[width=9cm]{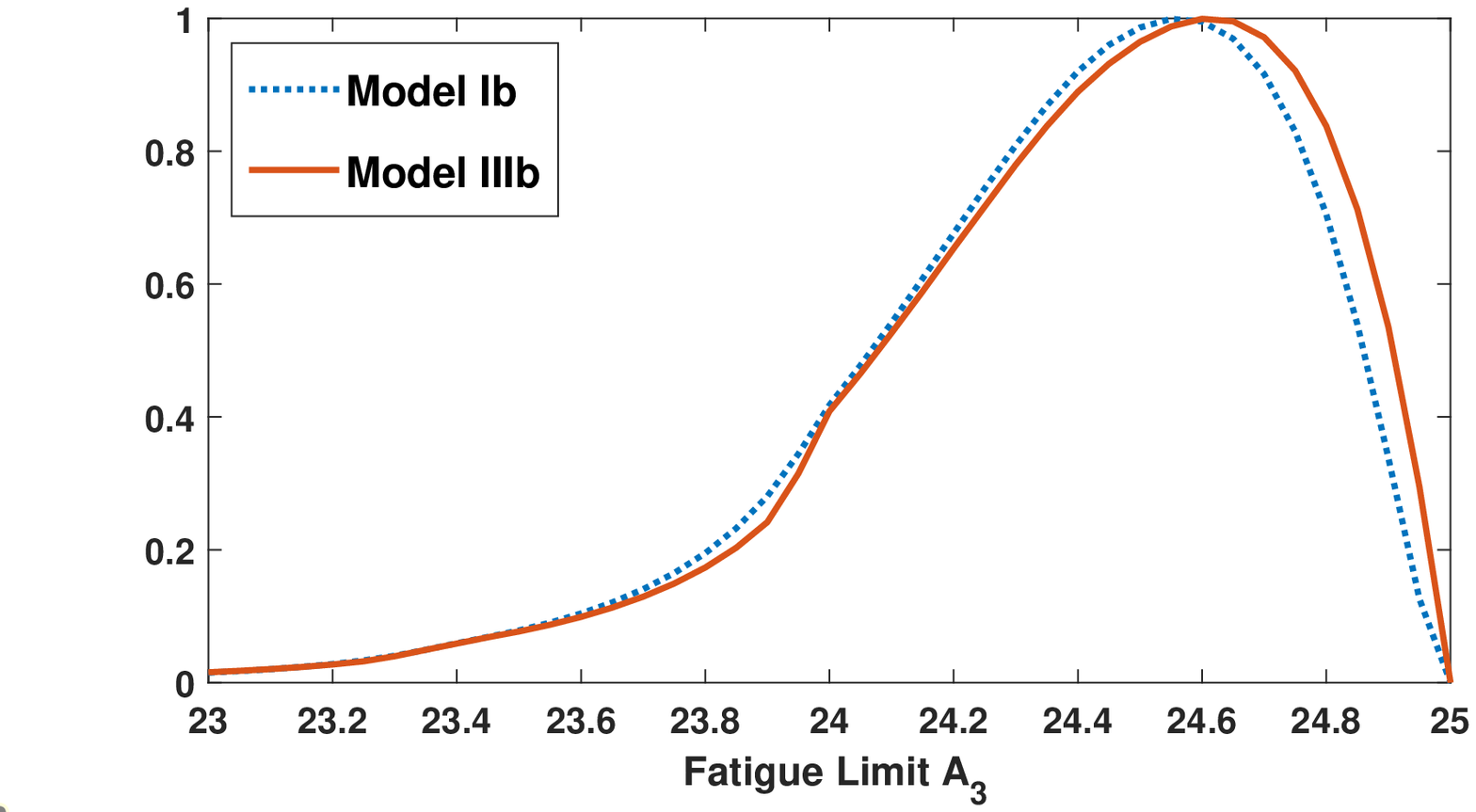}
  \captionof{figure}{Profile likelihoods of the fatigue-limit parameters using Models~Ib and IIIb.}
  \label{prof_like2}
\end{minipage}
% \caption{Profile likelihoods of the fatigue-limit parameters using Models Ia,IIa,Ib and IIb.}
\end{figure}

We compare the profile likelihood of the fatigue limit obtained using the previous models. Figure~\ref{prof_like1} depicts the profile likelihood of the fatigue limit, $A_3$, using Models~Ia and IIIa. For constant variance and shape parameters, the estimated profile likelihood using the Birnbaum--Saunders distribution (Model~IIIa) has a noticeably higher mode and a lower variance than Model~Ia, which uses the normal distribution. When adopting nonconstant variance and shape parameters, the difference between the two profile likelihoods is negligible, as displayed in Figure~\ref{prof_like2}.

\subsection{Survival functions}
\begin{figure}[h!]
\centering
\begin{minipage}{.45\textwidth}
  \centering
  \includegraphics[width=9cm]{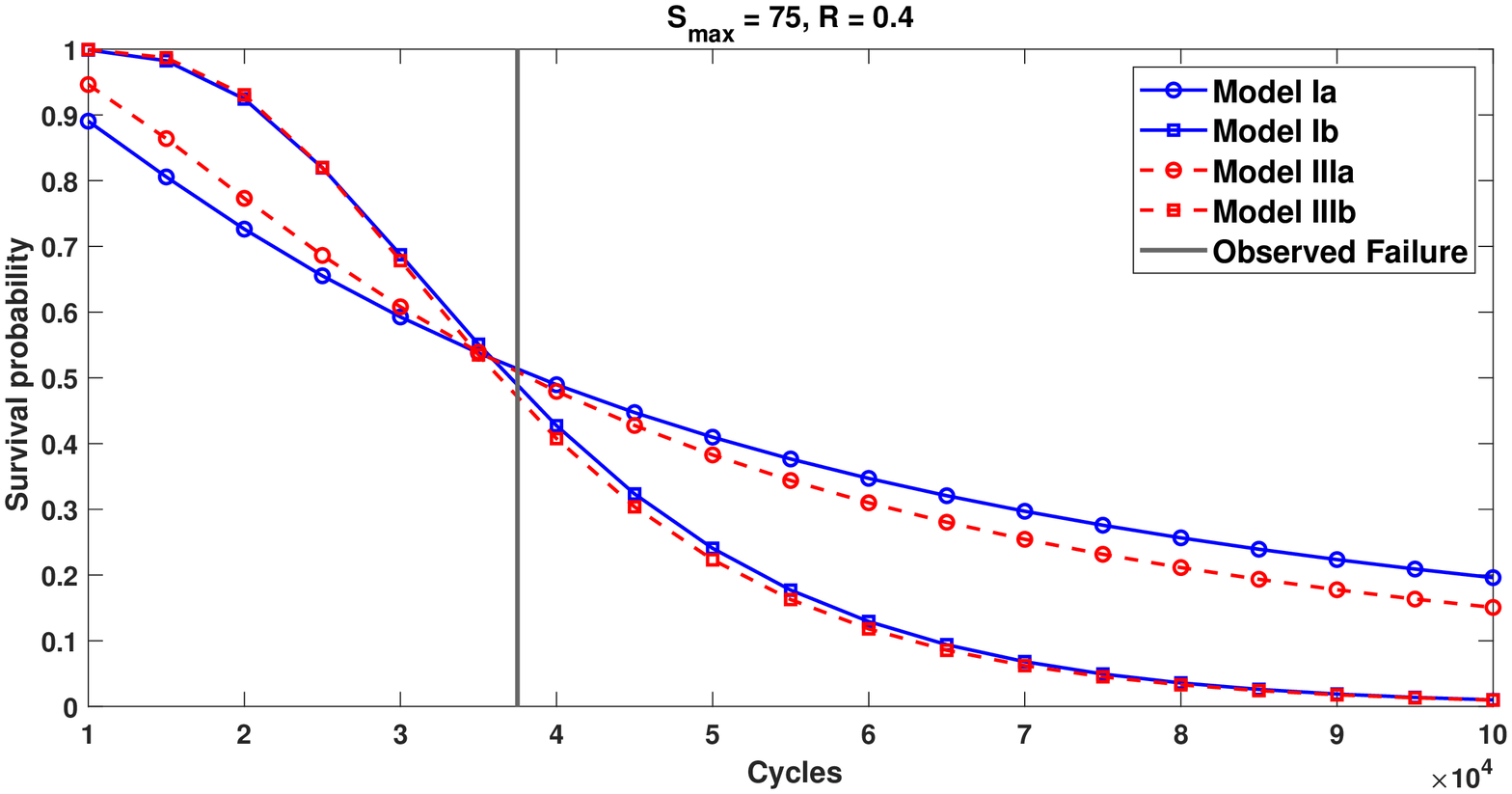}
  %\captionof{figure}{Survival functions of Dataset 1 specimens using the calibrated Models Ia, Ib, IIIa and IIIb when $S_{max} = 75$ and $R = 0.4$. }
  %\label{surv1}
\end{minipage}
~
\begin{minipage}{.45\textwidth}
\centering
  \includegraphics[width=9cm]{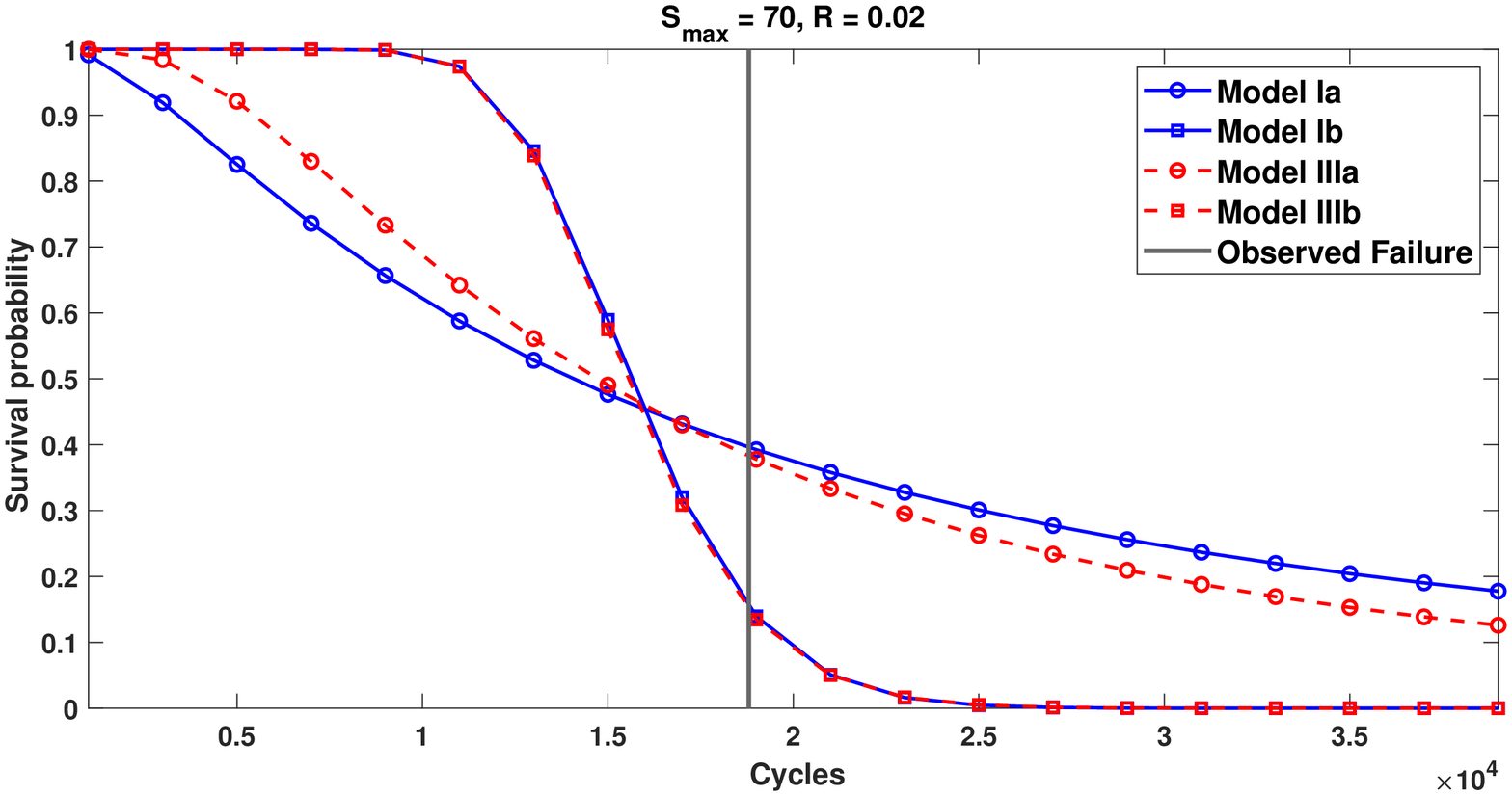}
  %\captionof{figure}{Survival functions of Dataset 1 specimens using the calibrated Models Ia, Ib, IIIa and IIIb when $S_{max} = 70$ and $R = 0.02$.}
  %\label{surv2}
\end{minipage}
~
\begin{minipage}{.45\textwidth}
\centering
  \includegraphics[width=9cm]{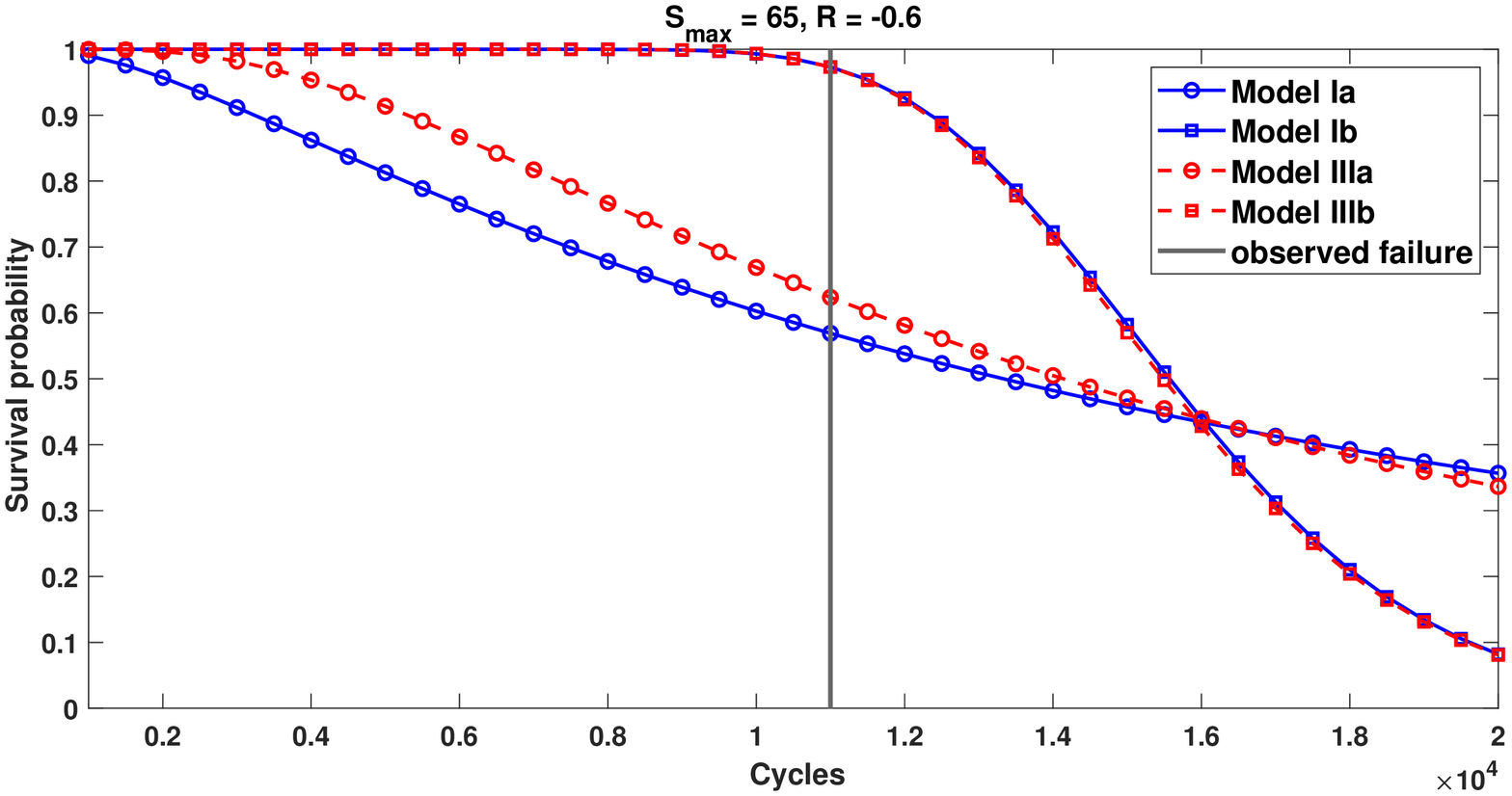}
  %\captionof{figure}{Survival functions of specimen \ref{specimen_data1} using the calibrated Models Ia, Ib, IIIa and IIIb when $S_{max} = 65$ and $R = -0.6$.}
  %\label{surv3}
\end{minipage}
~
\begin{minipage}{.45\textwidth}
\centering
  \includegraphics[width=9cm]{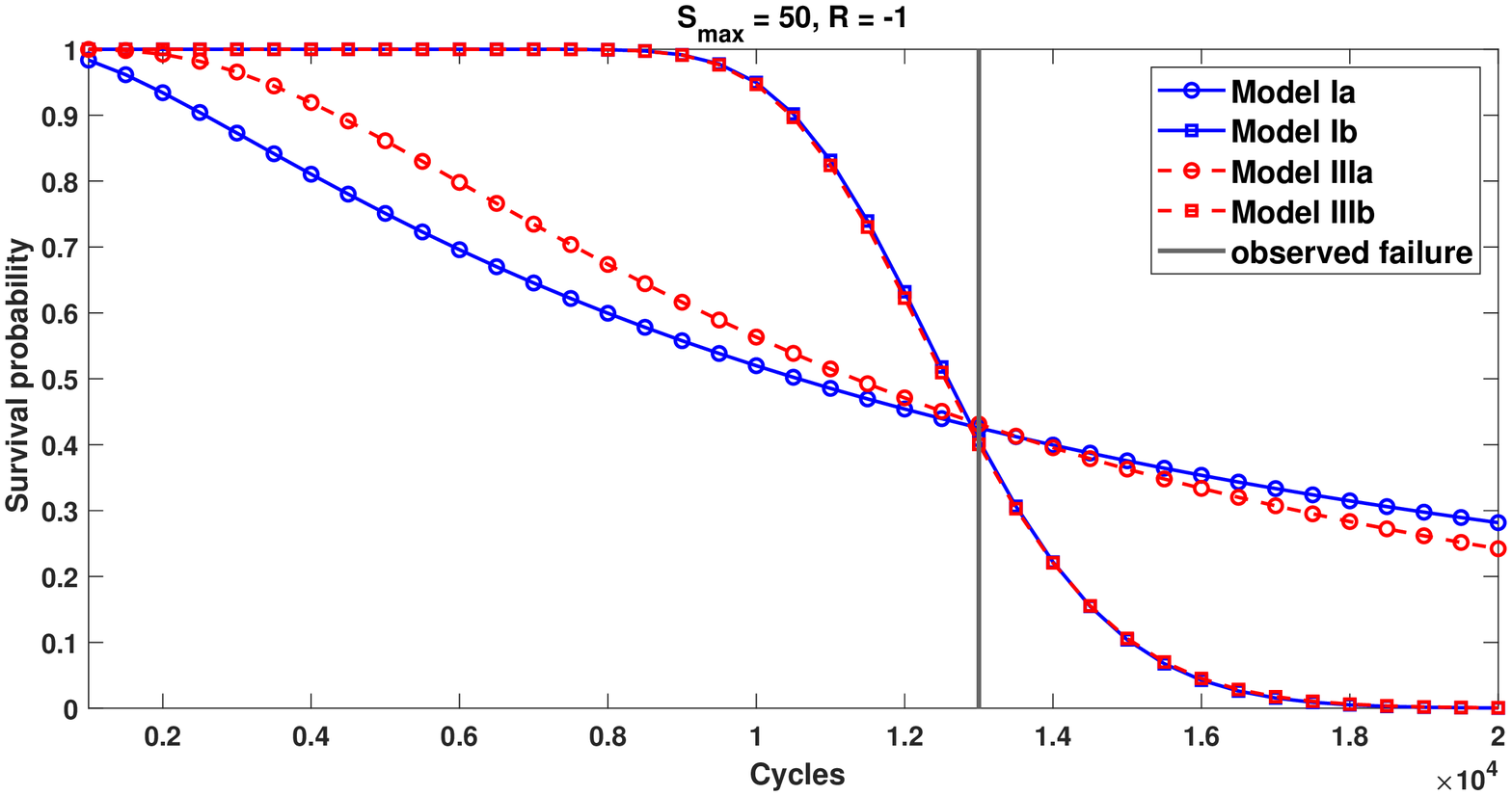}
  %\captionof{figure}{Survival functions of specimen \ref{specimen_data1} using the calibrated Models Ia, Ib, IIIa and IIIb when $S_{max} = 50$ and $R = -1$.}
  %\label{surv4}
\end{minipage}
\caption{Survival functions of Dataset 1 specimens using calibrated Models~Ia, Ib, IIIa, and IIIb for different values of $S_{max}$ and $R$.}
\label{surv_data1}
\end{figure}

We closely examined the survival functions obtained by calibrated Models~Ia, Ib, IIIa, and IIIb at different values of $S_{max}$ and $R$ in Figure~\ref{surv_data1}. The Birnbaum--Saunders model (Model IIIa) outperformed the counterpart Gaussian model (Model Ia) because it offers a higher survival probability before the observed failure and a lower survival probability after the observed failure. For Models~IIIa and IIIb, the resulting survival probabilities are almost identical for both distributions. 

\section{Model calibration and comparison for Dataset~2}
\label{sec4}

\subsection{Description of Dataset~2}
\label{sec4.1}
This section introduces and studies new datasets for unnotched specimens of 75S-T6 aluminum alloys \cite{hlg}. These datasets correspond to 101 round bar specimens with five minimum-section diameters. In Dataset~2, the fatigue experiments are rotating-bending, and the stress ratio is $-1$. Out of the 101 specimens, 13 experiments are run-outs.

% \subsection{Models }
% \label{sec4.2}
Again, we consider fatigue-limit Models~Ia, Ib, IIIa, and IIIb with the new equivalent stress \eqref{eqnew} to fit the data introduced in \ref{sec4.1}. As mentioned, the stress ratio for rotating-bending experiments is $-1$; therefore, the equivalent stress equals $S_{max}$.

\begin{table}[h!]
\begin{center}
\caption{Maximum likelihood estimates for Models~I and III with $S_{eq} = S_{max}$.}
\begin{tabular}{|c|c|c|c|c|c|c|c|c|}
\hline
Model & Specimen & Diameter & $A_1$ & $A_2$ & $A_3$ & $\tau / \alpha /  B_1$ &  $B_2$ & Max log-likelihood \\
\hline
Ia & 1 & 1/8  & 7.20 & -1.73 & 21.17 & 0.4420 & --- & -428.28  \\
\hline
Ib & 1 & 1/8 & 7.36 & -1.87 & 21.11 & 2.38 & -1.87 & -425.73  \\
\hline
IIIa & 1 & 1/8 & 7.17 & -1.72 & 21.23 & 0.0721 & --- & -426.06  \\
\hline
IIIb & 1 & 1/8 & 7.28 & -1.82 & 21.21 & 0.88 & -1.38 & -424.62 \\
\hline
Ia & 2 &  1/4 & 8.21 & -2.69 & 21.86 & 0.4002 & --- & -343.94 \\
\hline
Ib & 2 & 1/4 & 7.21 & -1.86 & 22.88 & 8.70 & -6.33 & -327.49 \\
\hline
IIIa & 2 & 1/4 & 7.96 & -2.48 & 22.16 & 0.0604 & --- & -340.75  \\
\hline
IIIb & 2 & 1/4 & 7.20 & -1.85 & 22.91 & 6.86 & -5.62 & -326.85 \\
\hline
Ia & All &  ---  & 9.03 & -3.04 & 18.68 & 0.5626 & --- & -1357.8  \\
\hline
Ib & All & --- & 8.23 & -2.50 & 20.49 & 3.39 & -2.53 & -1338.1  \\
\hline
IIIa & All & --- & 8.73 & -2.84 & 19.24 & 0.0880 & --- & -1349.6  \\
\hline
IIIb & All & --- & 8.15 & -2.45 & 20.61 & 1.86 & -2.02 & -1336.1 \\
\hline
\end{tabular}
\label{SND2}
\end{center}
\end{table}

\begin{figure}[h!]
\includegraphics[width=18cm]{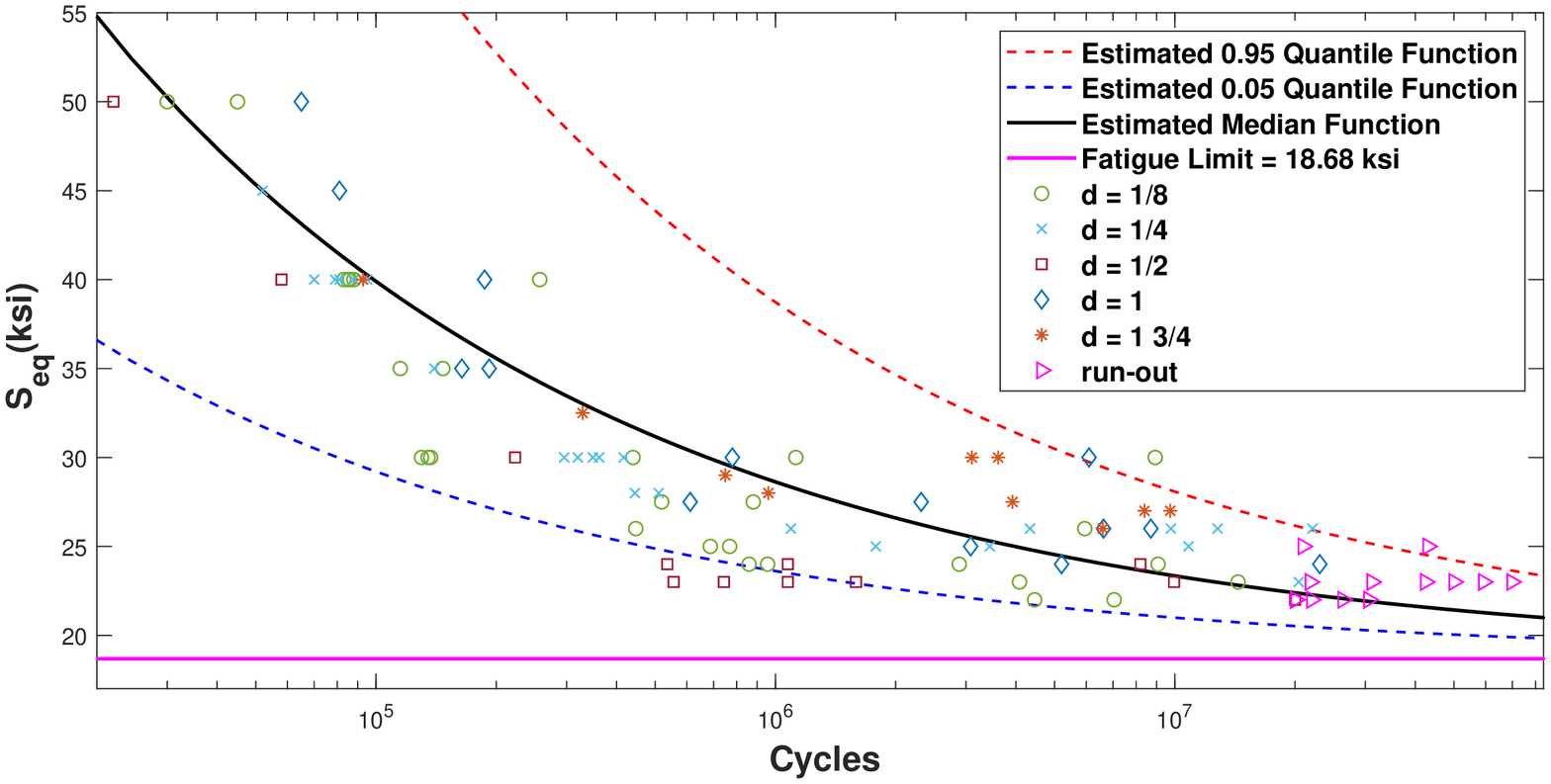}
\caption{Model Ia: $\log_{10}(N) \sim N(\mu(S_{eq}), \sigma)$ and $S_{eq} = S_{max}$.}
\label{fit1_data2}
\end{figure} 

\begin{figure}[h!]
\includegraphics[width=18cm]{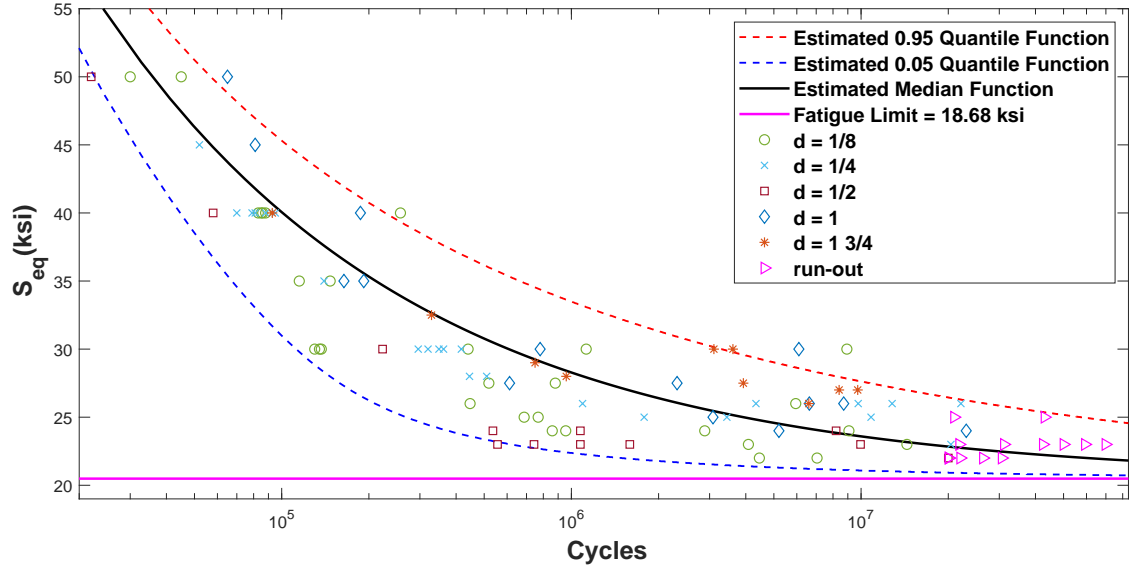}
\caption{Model Ib: $\log_{10}(N) \sim N(\alpha(S_{eq}), \mu(S_{eq}))$ and $S_{eq} = S_{max}$.}
\label{fit2_data2}
\end{figure}

Table~\ref{SND2} provides the MLEs for Models~Ia, Ib, IIIa, and IIIb when separately fitting Specimen~1 and 2. The joint fit for all specimens is also provided. The goodness of fit and estimated fatigue limit decreased when the data were combined. Figures~\ref{fit1_data2} and \ref{fit2_data2} reveal the quantiles of calibrated Models~Ia and Ib, respectively.  

\subsection{Profile likelihood and confidence intervals}
\begin{figure}[h!]
\centering
\begin{minipage}{.45\textwidth}
  \centering
  \includegraphics[width=9cm]{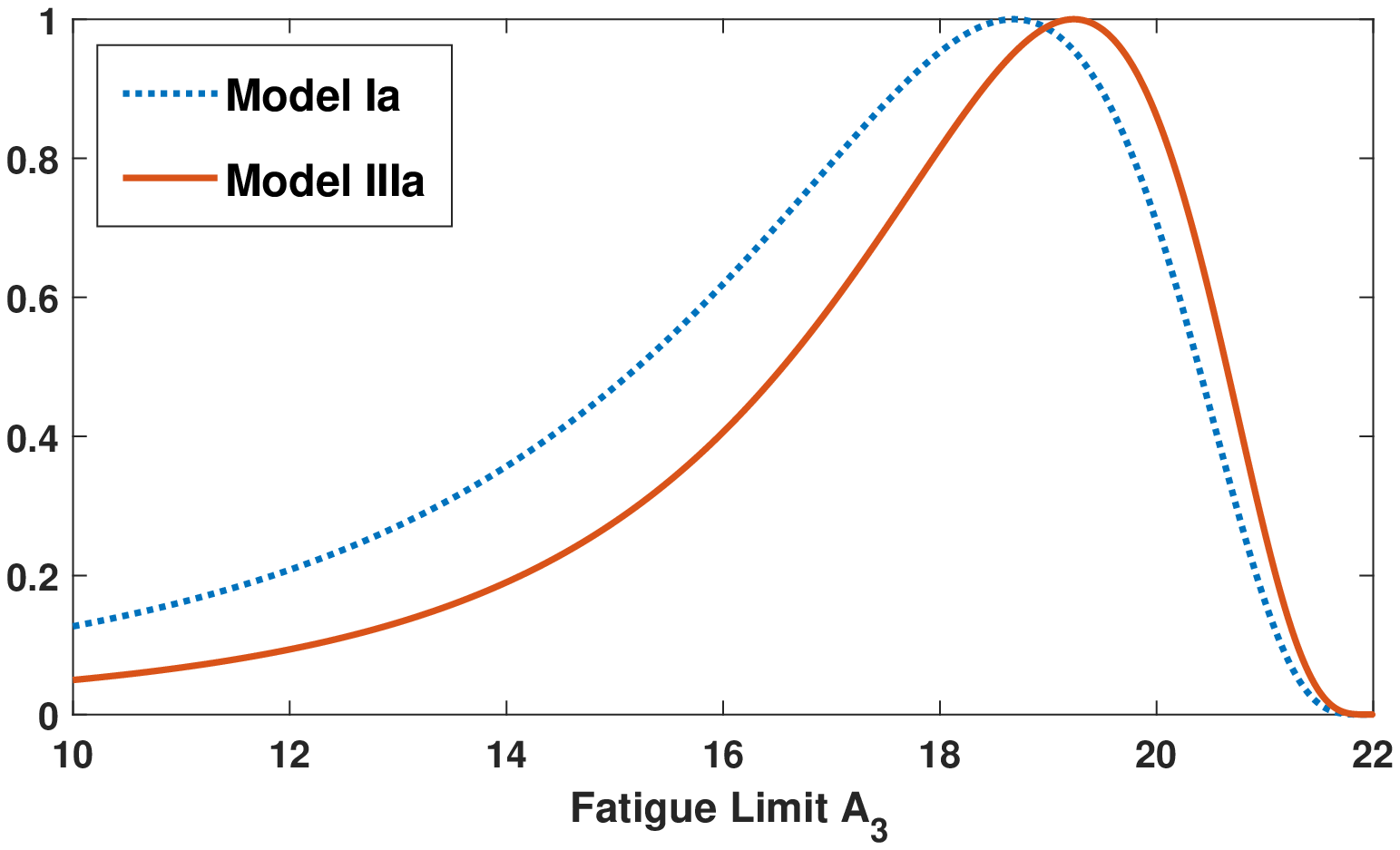}
  \captionof{figure}{Profile likelihoods of the fatigue-limit parameters using Models~Ia and IIIa.}
  \label{prof_like3}
\end{minipage}
~
\begin{minipage}{.45\textwidth}
\centering
  \includegraphics[width=9cm]{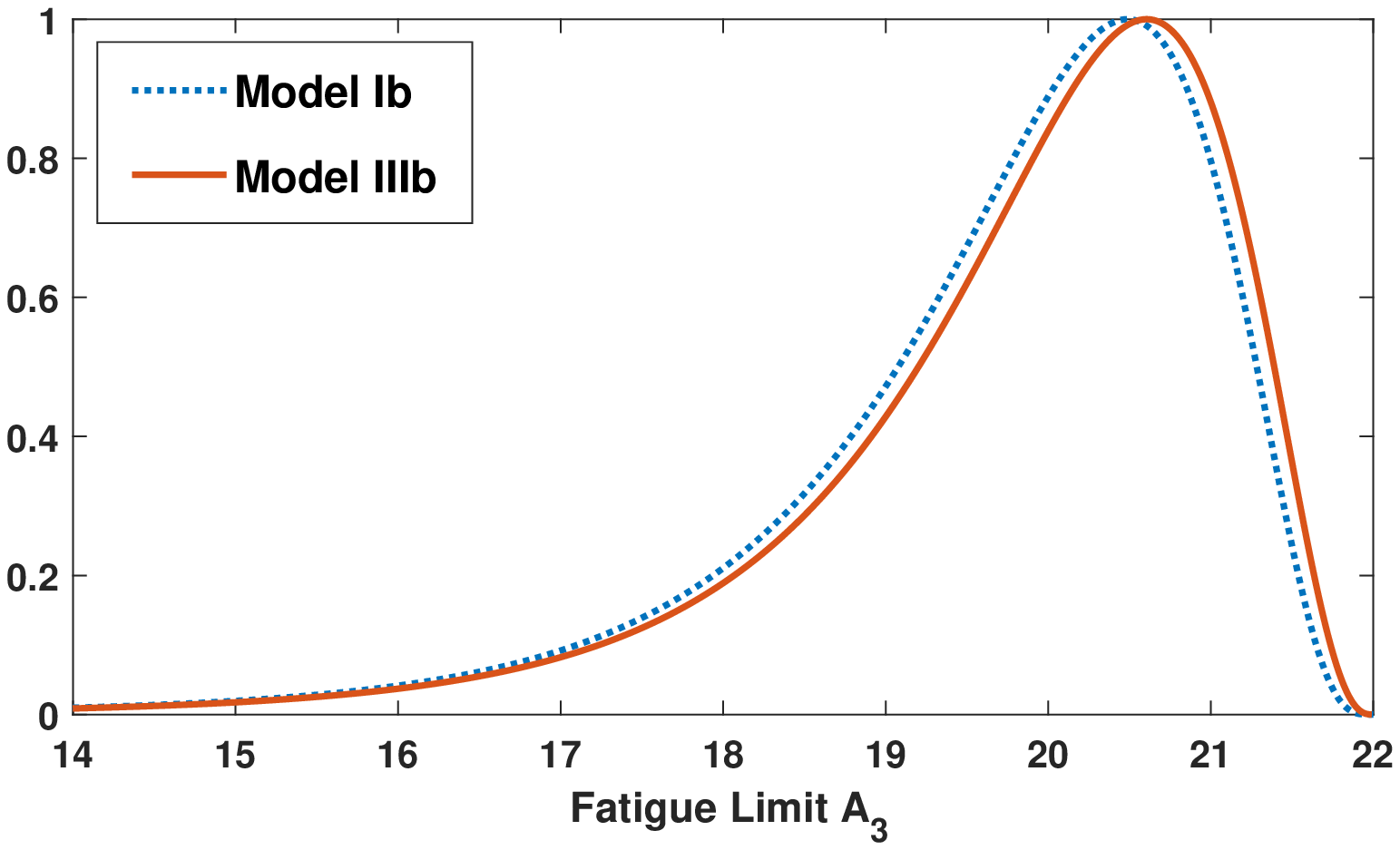}
  \captionof{figure}{Profile likelihoods of the fatigue-limit parameters using Models~Ib and IIIb.}
  \label{prof_like4}
\end{minipage}
% \caption{Profile likelihoods of the fatigue-limit parameters using Models Ia,IIa,Ib and IIb.}
\end{figure}

We again compare the profile likelihood of the fatigue limit obtained from the four previous models using Dataset~2. Figure~\ref{prof_like3} displays the profile likelihood of the fatigue limit, $A_3$, using Models~Ia and IIIa. As concluded, the profile likelihood when using the Birnbaum--Saunders distribution (Model IIIa) has a higher mode and much lower variance than Model Ia. With nonconstant variance and shape parameters, the two profile likelihoods are almost identical \ref{prof_like2}.

We confirmed the mentioned conclusions by estimating the confidence intervals of the pooled MLEs, given that Dataset~2 is complete. The confidence intervals presented in Table~\ref{CI_data2} are obtained by stratified bootstrapping where the sampled dataset maintains the same proportions in the original data related to the five specimens. The results indicate that the Birnbaum--Saunders distribution provides tighter confidence intervals than the normal distribution, especially when using a constant variance. This property is essential to generate accurate survival and failure predictions.

\begin{table}[h!]
\begin{center}
\caption{Confidence intervals of 90\% for the pooled maximum likelihood estimates for Models~I and III.}
\begin{tabular}{|c|c|c|c|c|c|c|c|}
\hline
Model & $A_1$ & $A_2$ & $A_3$ & $\tau / \alpha /  B_1$ &  $B_2$ \\
\hline
Ia & (7.9, 11.1) & (-4.3, -2.2) & (14.3, 21) & (0.49, 0.62) & --- \\
\hline
Ib & (7.6, 9.2) & (-3.1, -2.1) & (18.5, 21.6) & (2.8, 4.6) & (-3.4, -2.1)  \\
\hline
IIa & (7.8,10.3) & (-3.8,-2.2) & (15.7,21.1) & (0.077,0.096) & --- \\
\hline
IIb & (7.6, 9.1) & (-3.1,-2.0) & (18.6, 21.7) & (1.3, 3) & (-2.9,-1.6) \\
\hline
\end{tabular}
\label{CI_data2}
\end{center}
\end{table}

\subsection{Survival functions}
\begin{figure}[h!]
\centering
\begin{minipage}{.45\textwidth}
  \centering
  \includegraphics[width=9cm]{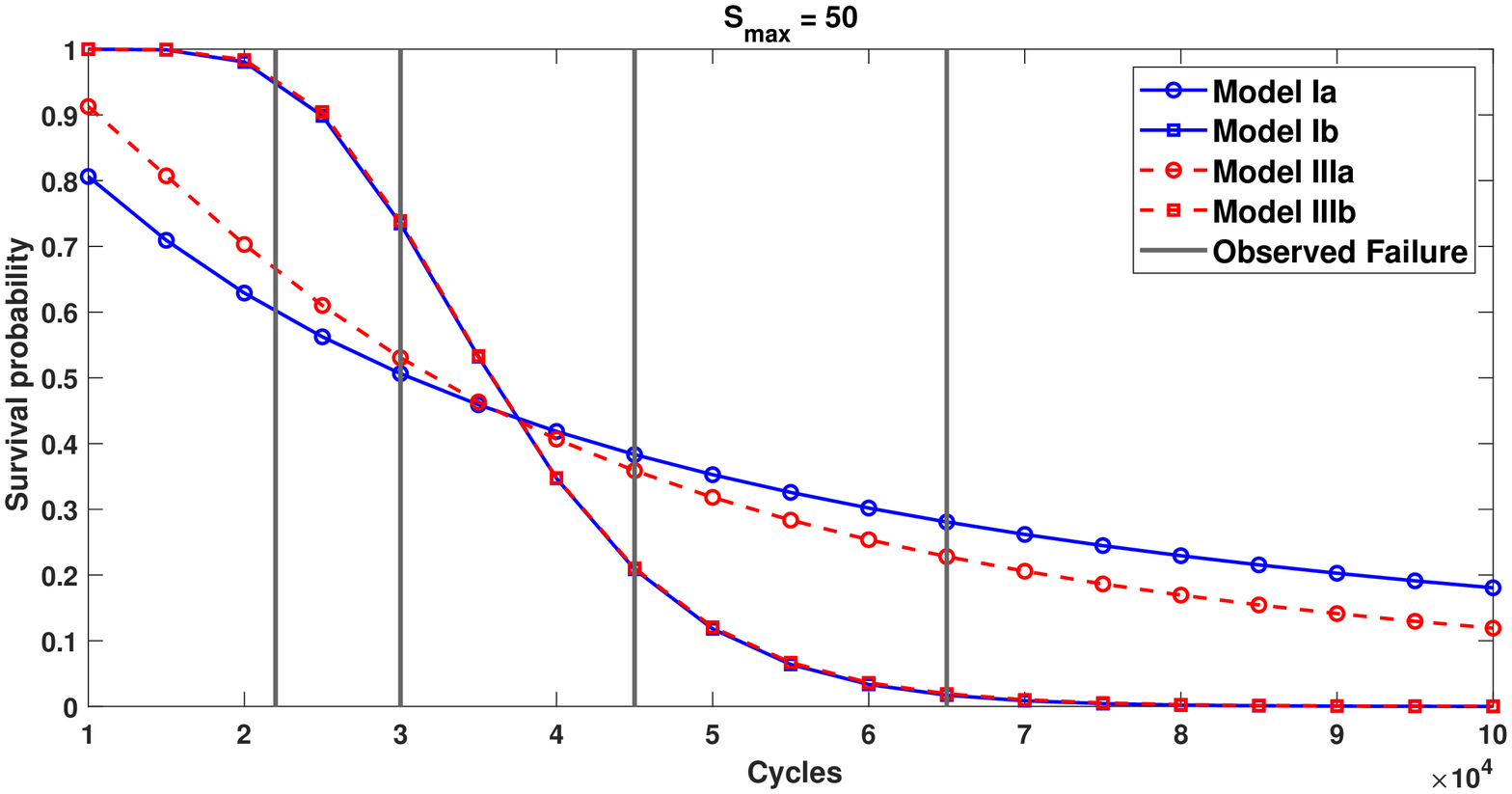}
  %\captionof{figure}{Survival functions of Dataset 1 specimens using the calibrated Models Ia, Ib, IIIa and IIIb when $S_{max} = 75$ and $R = 0.4$. }
  %\label{surv1}
\end{minipage}
~
\begin{minipage}{.45\textwidth}
\centering
  \includegraphics[width=9cm]{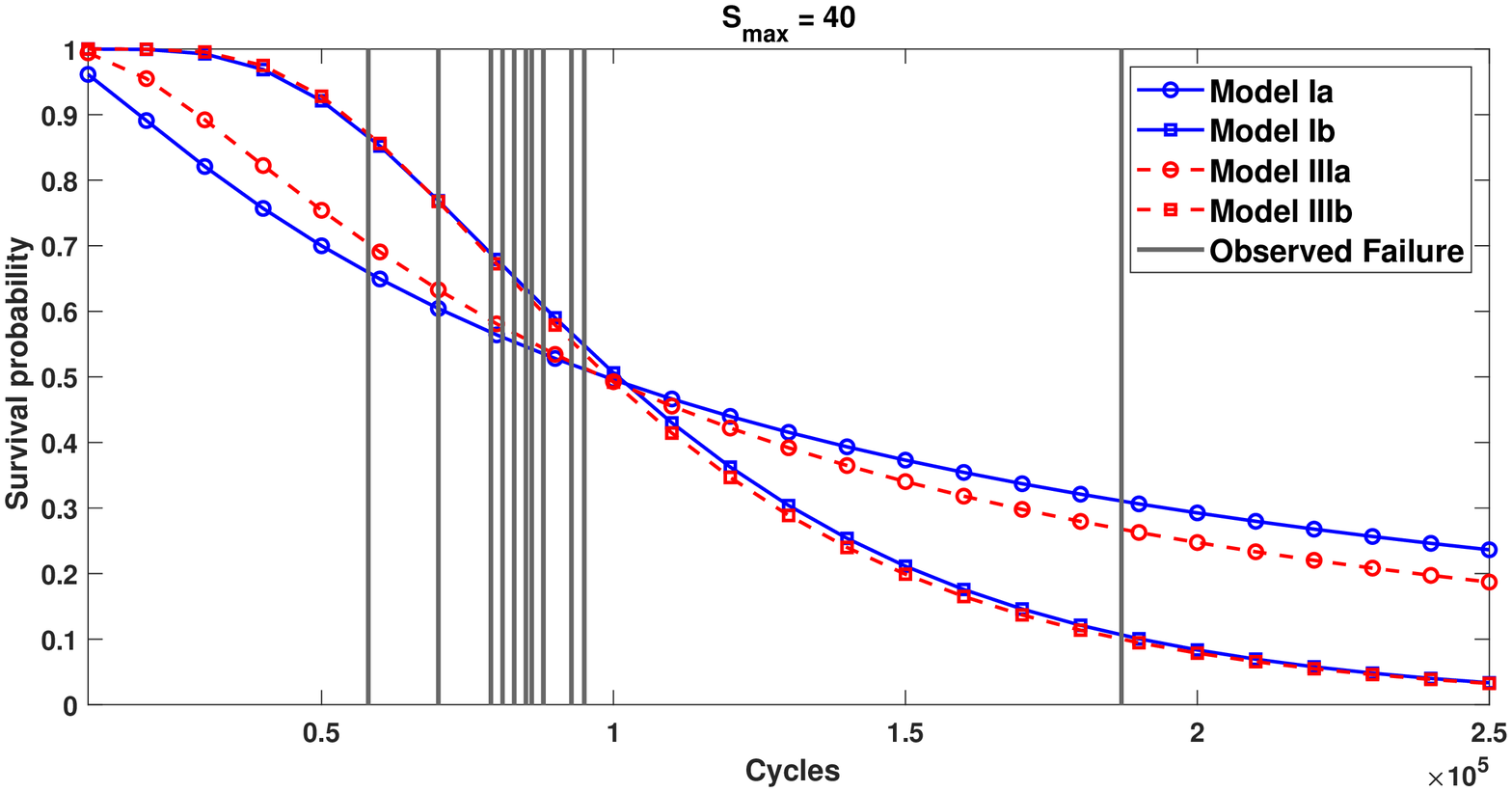}
  %\captionof{figure}{Survival functions of Dataset 1 specimens using calibrated Models Ia, Ib, IIIa, and IIIb when $S_{max} = 70$ and $R = 0.02$.}
  %\label{surv2}
\end{minipage}
\begin{minipage}{.45\textwidth}
\centering
  \includegraphics[width=9cm]{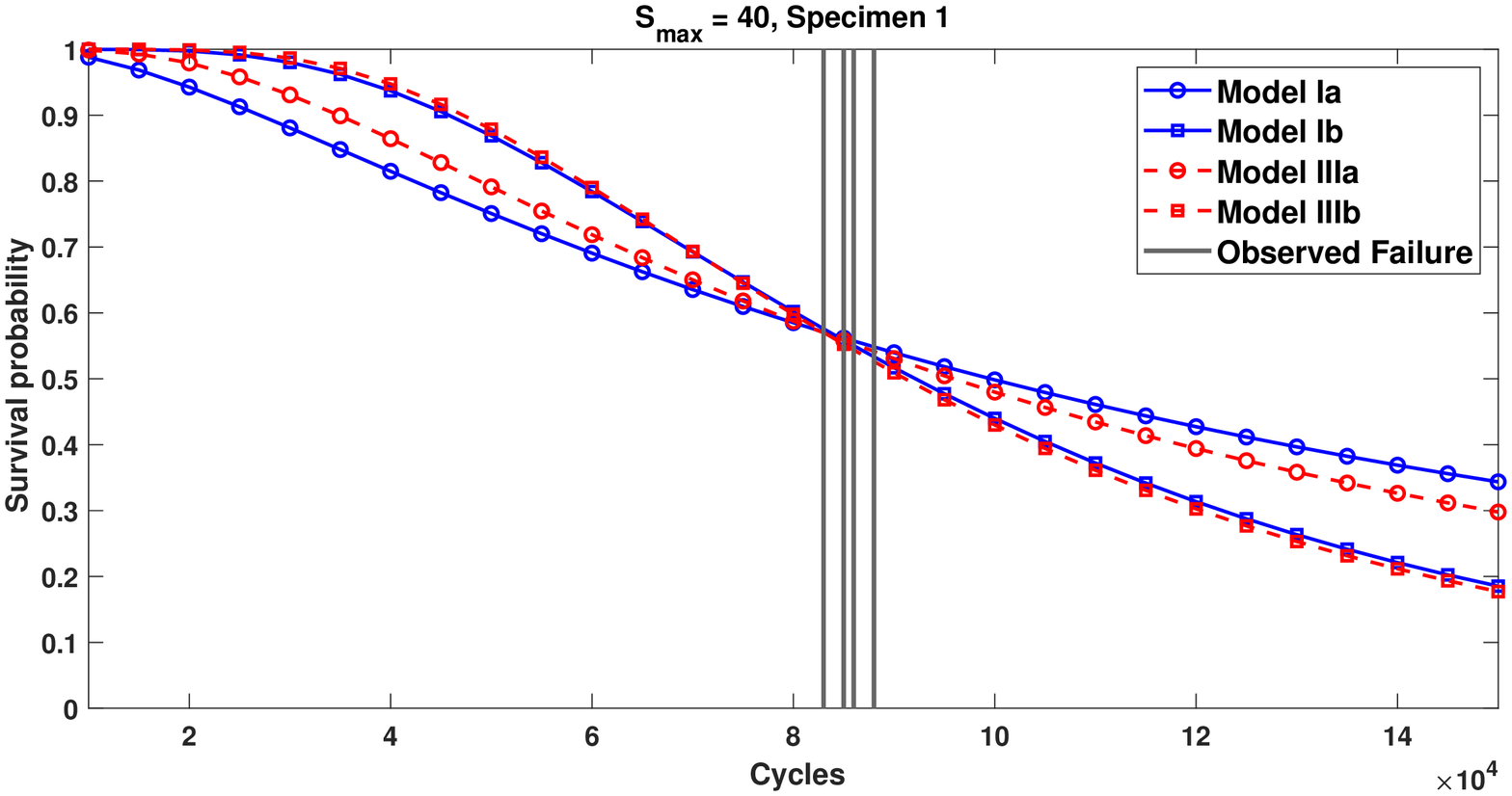}
  %\captionof{figure}{Survival functions of specimen \ref{specimen_data1} using calibrated Models Ia, Ib, IIIa, and IIIb when $S_{max} = 65$ and $R = -0.6$.}
  %\label{surv3}
\end{minipage}
~
\begin{minipage}{.45\textwidth}
\centering
  \includegraphics[width=9cm]{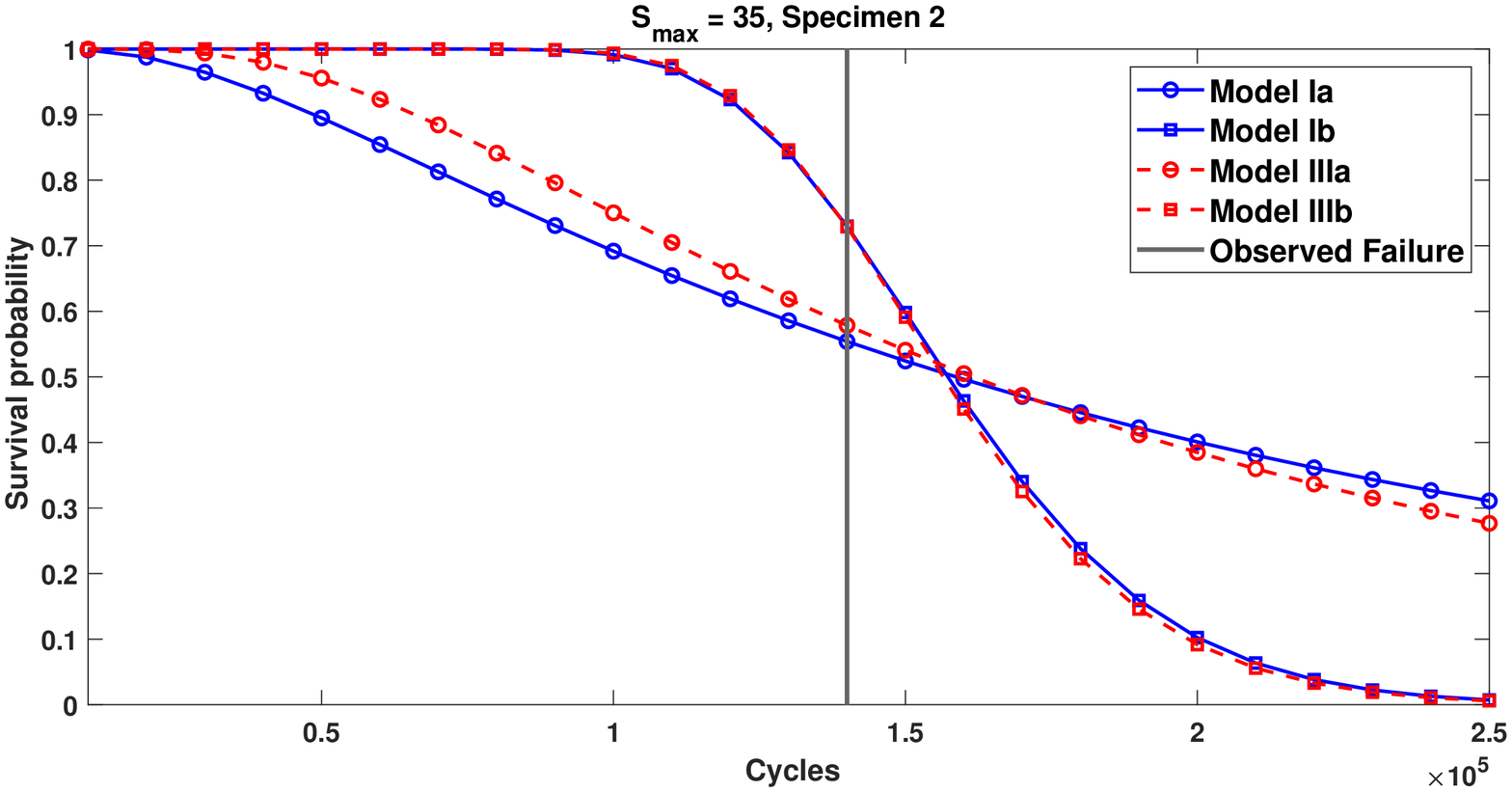}
  %\captionof{figure}{Survival functions of specimen \ref{specimen_data1} using calibrated Models Ia, Ib, IIIa, and IIIb when $S_{max} = 50$ and $R = -1$.}
  \label{surv4}
\end{minipage}
\caption{Survival functions of Dataset~2 specimens using calibrated Models~Ia, Ib, IIIa, and IIIb for values of $S_{max}$ and $R$.}
\label{surv_data2}
\end{figure}

Figure~\ref{surv_data2} depicts the survival probabilities of specimens from Dataset~2 under different settings using calibrated models by the pooled or specific specimen data. 

Comparing the results for Datasets~1 and 2, we notice higher variability in the latter, especially when using pooled Dataset~2 with multiple specimens of different geometries and sizes. The fit results could be improved using Poisson models that consider the geometry and size of the specimen \cite{fatigue2}. However, implementing and analyzing such models is beyond the scope of the current work.

\section{Model calibration and comparison for Dataset~3}
\label{sec5}

\subsection{Description of Dataset 3}
\label{sec5.1}
To generalize the previous results, we consider a well-known dataset: the laminate panel S-N dataset \cite{shimokawa1987, pasmee}. This dataset contains fatigue data for 125 carbon eight-harness-satin/epoxy laminate specimens subjected to four-point out-of-plane bending tests, where 10 out of 125 experiments are run-outs. In this case, the equivalent stress needed in the fatigue-limit models is given directly in the data, and we do not have the stress ratio.  

\begin{figure}[h!]
\centering
\includegraphics[width=16cm]{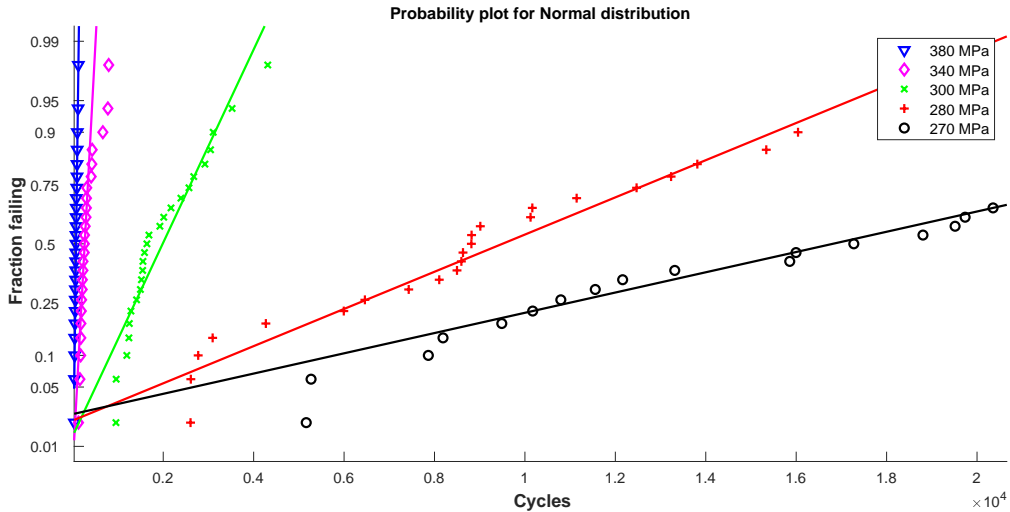}
\caption{Probability plot of the normal distribution as a model for the number of cycles $N$.}
\label{pp_normal}
\end{figure}

\begin{figure}[h!]
\centering
\includegraphics[width=16cm]{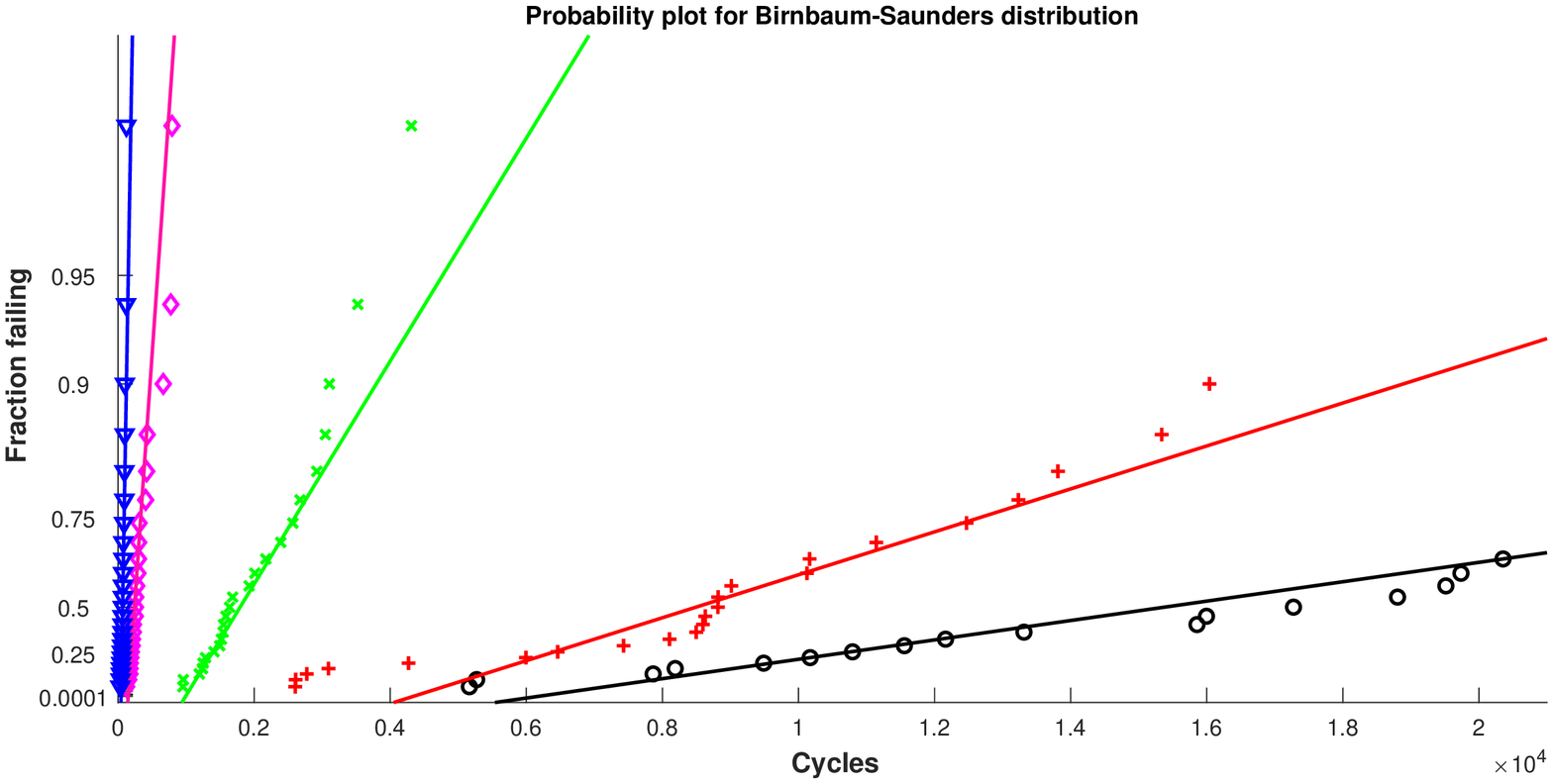}
\caption{Probability plot of the Birnbaum--Saunders distribution as a model for the number of cycles $N$.}
\label{pp_BS}
\end{figure}

\begin{figure}[h!]
\centering
\includegraphics[width=16cm]{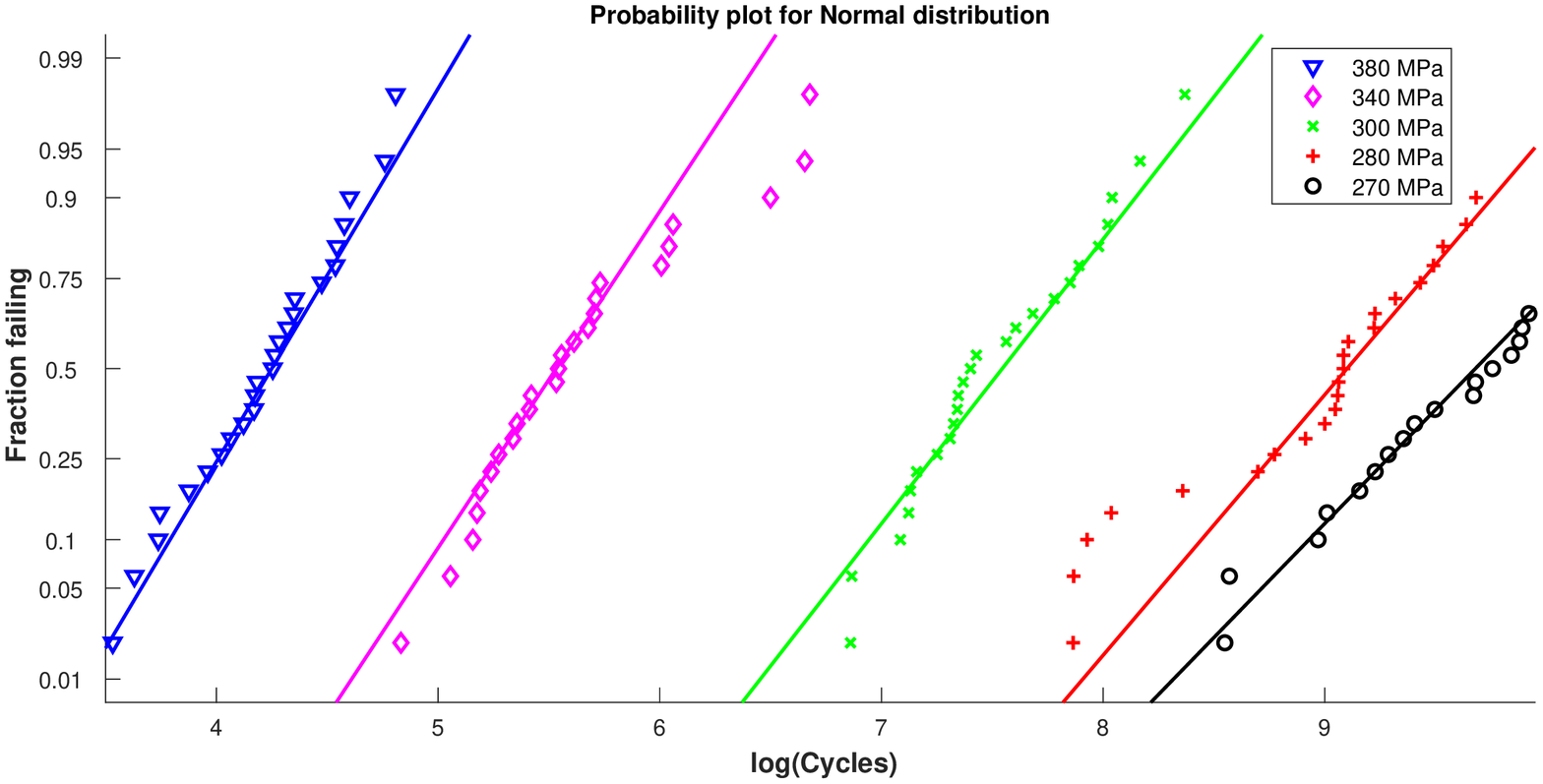}
\caption{Probability plot of the normal distribution as a model for $\log(N)$.}
\label{log_pp_normal}
\end{figure}

\begin{figure}[h!]
\centering
\includegraphics[width=16cm]{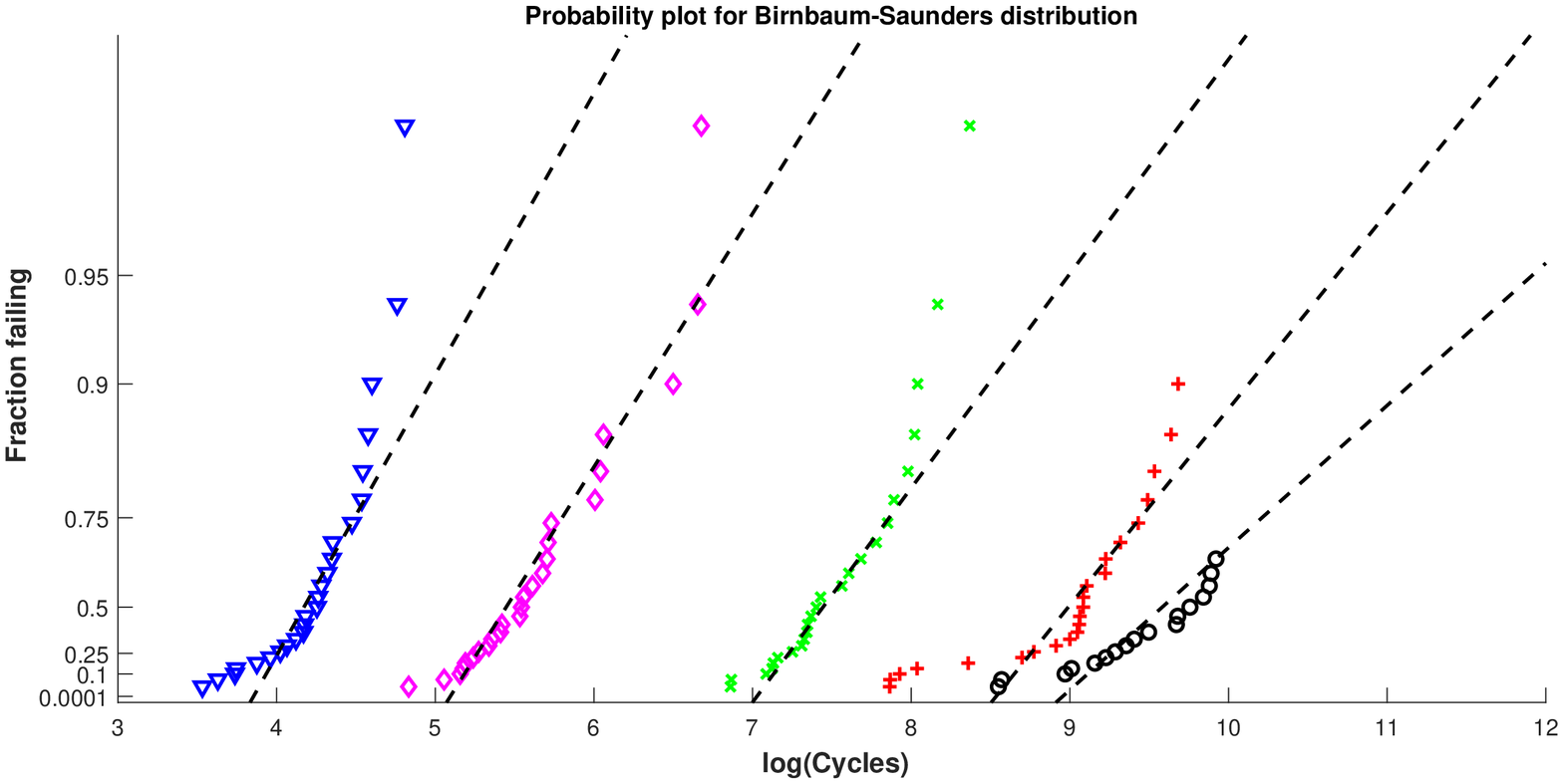}
\caption{Probability plot of the Birnbaum--Saunders distribution as a model for $\log(N)$.}
\label{log_pp_BS}
\end{figure}

As a first illustration, we used probability plots, as suggested in \cite{meeker2022modern}. Figures~\ref{pp_normal} and \ref{pp_BS} present the probability plot of the normal distribution and Birnbaum--Saunders distribution as models for the fatigue life, $N$. Modeling $N$ using the normal or Birnbaum--Saunders distribution is not a good choice. Instead, modeling $\log(N)$ using these distributions provides better probability plots, as illustrated in Figures~\ref{log_pp_normal} and \ref{log_pp_BS}.

We calibrated six fatigue-limit models: Ia, Ib, IIa, IIb, IIIa, and IIIb, slightly modified using natural instead of base 10 logarithms. This approach was conducted to make the MLE parameters comparable to the results in the literature and does not affect the goodness of fit. We also fit the data using the Weibull distribution but did not include these results, as this distribution consistently provides the worst fit.

\begin{table}[h!]
\begin{center}
\caption{Maximum likelihood estimates for Models~I, II, and III.}
\begin{tabular}{|c|c|c|c|c|c|c|c|c|}
\hline
Model & $A_1$ & $A_2$ & $A_3$ & $\tau / \alpha /  B_1$ &  $B_2$ & Max log-likelihood \\
\hline
Ia & 31.56 &  -5.32  &  209.69 &  0.4902 & --- & -889.77  \\
\hline
Ib &  30.26 & -5.10  &  214.22 &  8.71 & -1.64 & -885.28  \\
\hline
IIa & 32.15 & -5.43 & 207.55 & 0.5031 & --- & -889.90  \\
\hline
IIb & 30.77 & -5.18 & 212.45 &  9.01 & -1.69  & -885.17 \\
\hline
IIIa & 29.63  & -4.99  &  216.48 & 0.0718 & --- & -885.64  \\
\hline
IIIb & 30.14 & -5.08 & 214.59 &  -6.84 & 0.73  & -884.67 \\
\hline
\end{tabular}
\label{SND3}
\end{center}
\end{table}

\begin{table}[h!]
\begin{center}
\caption{Classical information criteria.}
\begin{tabular}{|c|c|c|c|c|c|c|c|}
\hline
Models &  \bf{Ia} & \bf{Ib} & \bf{IIa} & \bf{IIb} & \bf{IIIa} & \bf{IIIb} \\
\hline
Maximum log-likelihood & -889.77 & -885.28 & -889.90  &  -885.17  & -885.64 & -884.67 \\
\hline
Akaike information criterion (AIC) & 1787.5 & 1780.6 &  1787.8 &  1780.3  & 1779.3  & 1779.3 \\
\hline
Bayesian information criterion (BIC) & 1798.9 &  1794.7 & 1799.1  &  1794.5  &  1790.6 &  1793.5 \\
\hline
Akaike information criterion with correction & 1787.9  &  1781.1  &  1788.1 & 1780.8 &  1779.6  &  1779.8  \\
\hline
\end{tabular}
\label{CIC3}
\end{center}
\end{table}

\begin{figure}[h!]
\includegraphics[width=16cm]{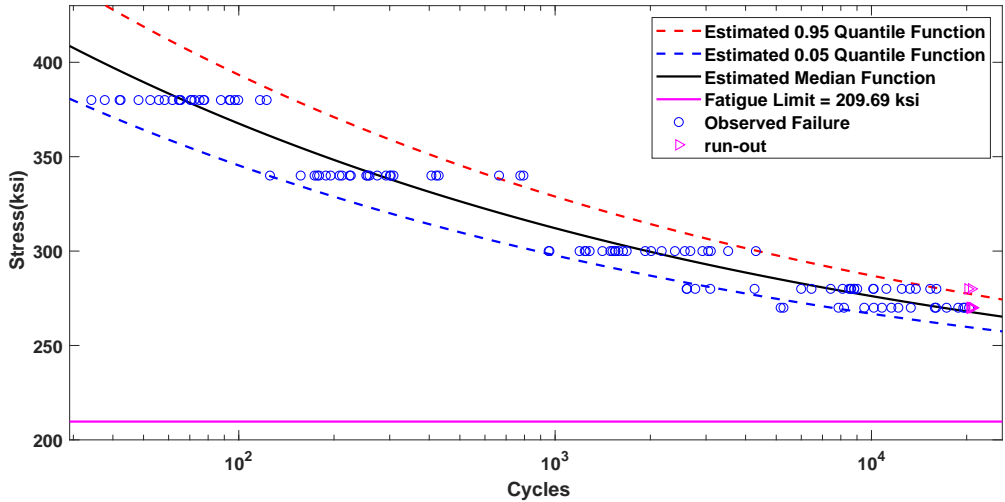}
\caption{Model Ia: $\log(N) \sim N(\mu(S), \sigma)$.}
\label{fit1_LN_data3}
\end{figure} 

\begin{figure}[h!]
\includegraphics[width=16cm]{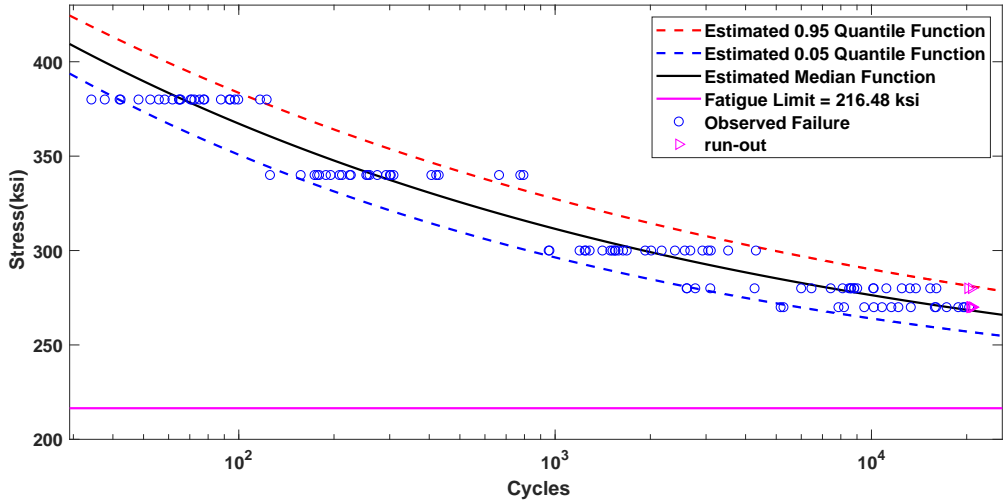}
\caption{Model IIIa: $\log(N) \sim BS(\alpha, \mu(S))$.}
\label{fit1_BS_data3}
\end{figure}

\begin{figure}[h!]
\includegraphics[width=16cm]{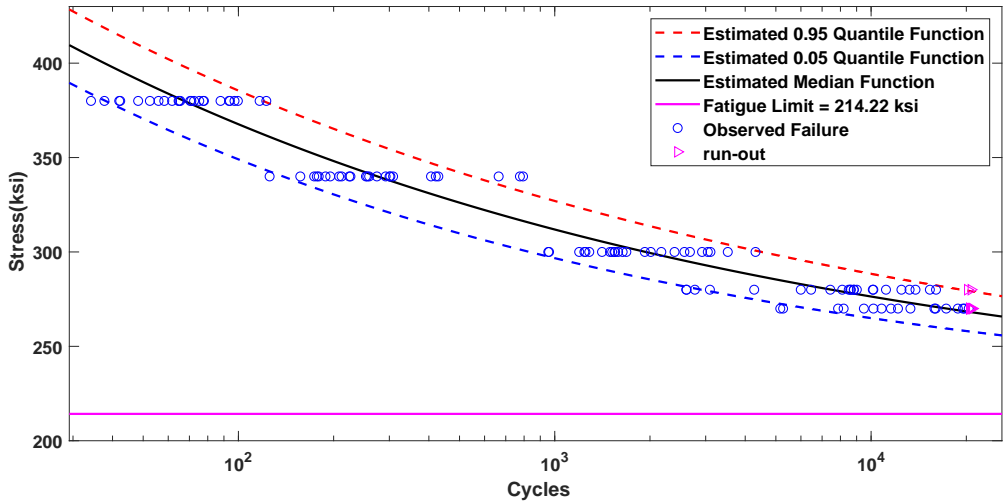}
\caption{Model Ib: $\log(N) \sim N(\mu(S), \sigma(S))$.}
\label{fit2_LN_data3}
\end{figure} 

\begin{figure}[h!]
\includegraphics[width=16cm]{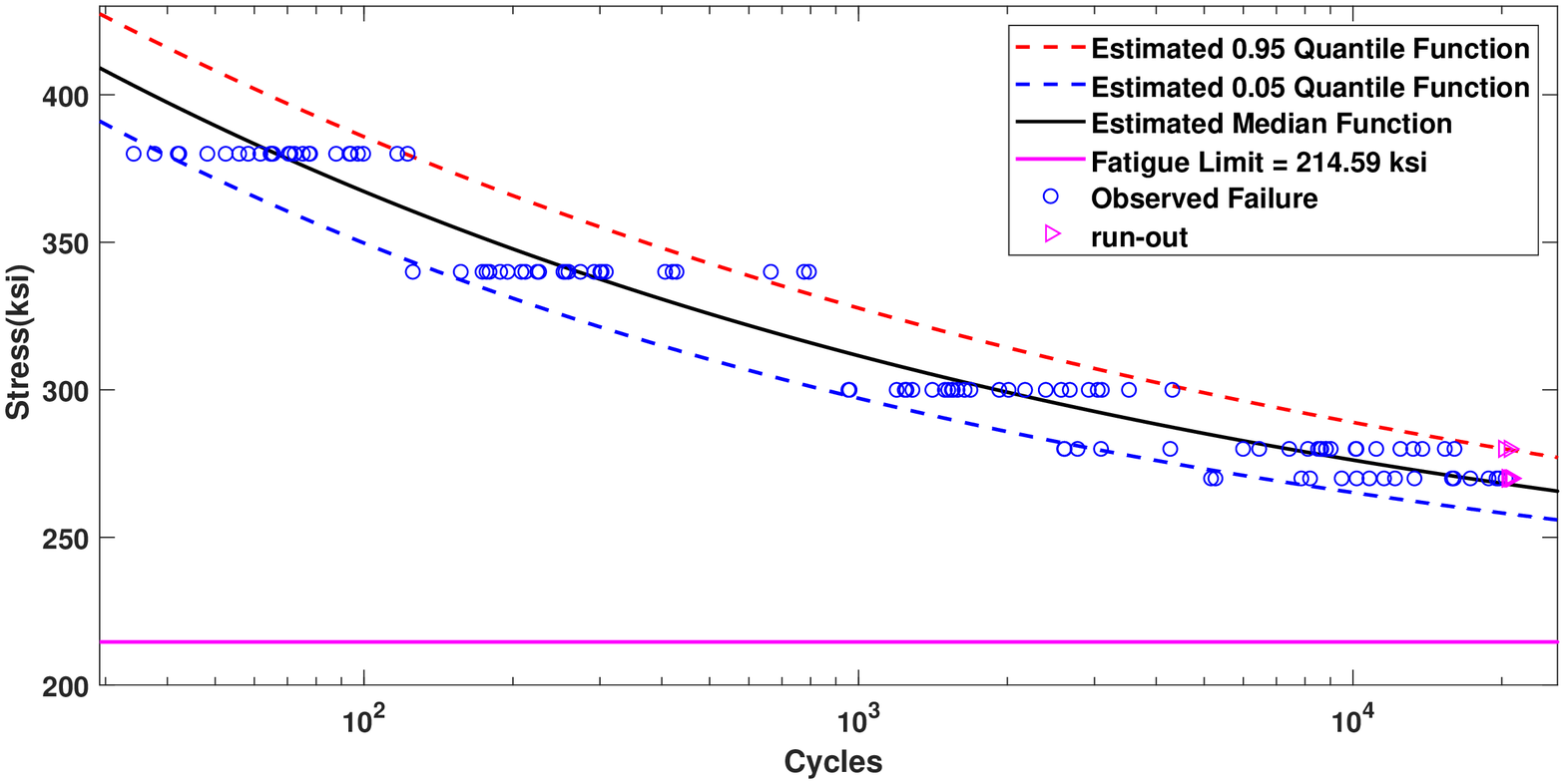}
\caption{Model IIIb: $\log(N) \sim BS(\alpha(S), \mu(S))$.}
\label{fit2_BS_data3}
\end{figure}

Table~\ref{SND3} provides the MLEs for Models~Ia, Ib, IIa, IIb, IIIa, and IIIb. Table~\ref{CIC3} compares all six models employing classical information criteria. Figures~\ref{fit1_LN_data3} and \ref{fit1_BS_data3} present the quantiles of calibrated Models~Ia and IIIa, respectively. Figures \ref{fit2_LN_data3} and \ref{fit2_BS_data3} display the quantiles of calibrated Models~Ib and IIIb, respectively. 

\section{Conclusions}
\label{sec6}
Multiple variants of the fatigue-limit models were calibrated and ranked employing ML and classical information criteria. The proposed approach of modeling the logarithm of the fatigue life using the Birnbaum--Saunders distribution proved to be superior or equivalent to the best model in all cases. 

For Dataset~1, fatigue experiments included two types of loadings. We introduced a new equivalent stress to generalize the models for such scenarios. Therefore, the fit for Dataset~1 considerably improved for all models. The suggested equivalent stress does not require adding new parameters and could be used for fatigue data with only one loading type.

In Dataset~2, five types of round bar specimens were subjected to rotating-bending fatigue experiments. The calibration was performed using Specimens~1 and 2 individually. Then, pooled calibration was performed using the full dataset. The variability of the latter calibration increased, which was further analyzed by obtaining confidence intervals of the MLEs via bootstrapping.

Laminate panel data were adopted in Section~\ref{sec5}. Models~I, II, and III were calibrated and ranked using constant and nonconstant variance/shape parameters. The fit results confirmed that Birnbaum--Saunders models are better than log-normal and Weibull models, especially when the variance is constant. Model~IIIa is preferable over IIIb when comparing the classical information criteria.

For all calibrated models, we analyzed the data with the estimated S-N curves and survival probabilities in the three datasets. The various models were compared using AIC, BIC, and AIC with correction. Profile likelihoods were also computed for the fatigue-limit parameter.

The Birnbaum--Saunders distribution provided a better fit for data and higher confidence in estimating the fatigue-limit parameter in Model~IIIa. Models~Ib and IIIb yielded similar results regarding information criteria, survival probabilities, and profile likelihood. A nonconstant variance with the log-normal distribution could be an alternative to the proposed Birnbaum--Saunders model in some frameworks.

% \begin{itemize}
%     \item The use of Birnbaum--Saunders distribution with the fatigue-limit models provides, in general, better fit than the normal distribution.
%     \item The equivalent stress (\ref{eqnew}) (i.e. $S_{eq} = S_{max}(\frac{1-R}{2})^{1-\sign(R)q}$) improves the fitting for Dataset 1.
%     \item There are five different types of specimen in Dataset 2. The results in Table \ref{SND2} show that the estimated parameters varies with the specimen type and the goodness of fit is reduced when the whole data is calibrated. %\textcolor{red}{Is this due to the size effect?}
%     %\item In Table \ref{PM1}, we use the Poisson model assuming the stress is almost uniform in the critical region $C$. However, the ratio $\frac{\gamma}{|C|}$ is estimated to be greater than 1. \textcolor{red}{How to explain that?}
%     %\item \textcolor{red}{Can we calibrate Dataset 1 and Dataset 2 jointly given that these data correspond to unnotched specimens?}
%     %\item \textcolor{red}{Although Dataset 1 and 2 correspond to the same material, the estimated fatigue-limit is different.}  
% \end{itemize}

%\section*{Acknowledgement}
%Z. Sawlan, M. Scavino and R. Tempone are members of King Abdullah University of Science and Technology (KAUST) SRI Center for Uncertainty Quantification in Computational Science and Engineering.
%The research reported in this publication was supported by King Abdullah University of Science and Technology (KAUST) through support from the Strategic Research Initiative on Uncertainty Quantification to Z. Sawlan, M. Scavino and R. Tempone.
%\section*{References}

\bibliography{mybib}

\end{document}